\newif\ifarxiv
\arxivtrue

\newif\ifcomm
\commfalse

\newif\ifacm
\acmfalse

\newif\ifusenix
\ifacm
   \usenixfalse %
\else
    \usenixfalse %
\fi

\newif\ifhotnets
\ifacm
   \hotnetsfalse %
\else
    \ifusenix
        \hotnetsfalse %
    \else
        \hotnetsfalse %
    \fi
\fi
    
\newif\ifacmart
\ifacm   %
   \acmarttrue
\else
   \acmartfalse
\fi

\newif\ifieee
\ifacm
   \ieeefalse
\else
    \ifusenix
        \ieeefalse
    \else
        \ifhotnets
            \ieeefalse
        \else
            \ieeetrue
        \fi
    \fi
\fi

\documentclass[journal]{IEEEtran}

\usepackage{amsmath,amsfonts}
\usepackage{algorithm}
\usepackage{array}
\usepackage{textcomp}
\usepackage{stfloats}
\usepackage{url}
\usepackage{verbatim}
\usepackage{graphicx}
\usepackage{cite}
\usepackage[hidelinks]{hyperref}
\hypersetup{pdftitle={Scheduling Parallel Optical Circuit Switches for AI Training},
            pdfauthor={Kevin Liang, Litao Qiao, Isaac Keslassy, Bill Lin}}

	\usepackage{amsthm}  
    \newtheoremstyle{boldthm}{}{}{\itshape}{}{\bfseries}{.}{ }{\thmname{#1}\thmnumber{ #2}\thmnote{ (#3)}} 
    
    \theoremstyle{boldthm}

  \newtheorem{theorem}{Theorem}\newtheorem{proposition}{Proposition}\newtheorem{property}{Property} \newtheorem{lemma}{Lemma}

\usepackage[justification=justified, font=footnotesize]{caption}
\usepackage[labelformat=simple, font=footnotesize]{subcaption}  
\DeclareCaptionLabelSeparator{periodspace}{.\quad}
\usepackage{graphicx}
\usepackage{amssymb}
\usepackage{xurl}
\usepackage[normalem]{ulem} \usepackage{xcolor} 

\usepackage{xspace}
\usepackage{balance}
\usepackage{booktabs}  
\usepackage{array}
	\newcolumntype{C}[1]{>{\centering\let\newline\\\arraybackslash\hspace{0pt}}m{#1}}
\usepackage[inline,shortlabels]{enumitem} 

\usepackage[capitalize,noabbrev]{cleveref}
\crefname{equation}{Eq.}{Eqs.}
\Crefname{equation}{Eq.}{Eqs.}
\crefname{figure}{Fig.}{Figs.}
\Crefname{figure}{Fig.}{Figs.}
\crefname{table}{Tab.}{Tabs.}
\Crefname{table}{Tab.}{Tabs.}
\crefname{algorithm}{Alg.}{Algorithms.}
\Crefname{algorithm}{Alg.}{Algorithms.}
\crefformat{section}{\S#2#1#3}
\crefformat{subsection}{\S#2#1#3}
\Crefformat{section}{\S\S#2#1#3} \Crefformat{subsection}{\S\S#2#1#3}

\newcommand{\SmallReduceVSpace}{}	
\newcommand{\ReduceVSpace}{}

\usepackage[inline]{enumitem}
\newcommand{\T}[1]{\par\smallskip\noindent\textbf{#1}} \newcommand{\Ts}[1]{\par\smallskip\noindent\textit{#1}}

\newcommand{\be}{\begin{equation}}
\newcommand{\ee}{\end{equation}}

\newcommand{\para}[1]{\left( #1 \right)}        

\newcommand{\set}[1]{\left\{#1\right\}}         
\newcommand{\sbrac}[1]{\left[ #1 \right]}

\newcommand{\unit}[1]{\,\mathrm{#1}} 

\DeclareMathOperator*{\argmin}{argmin} \DeclareMathOperator*{\argmax}{argmax}

\newcommand{\vx}{\checkmark\kern-1.1ex\raisebox{.7ex}{\rotatebox[origin=c]{125}{--}}}

\newcommand{\newVar}[2]{\newcommand{#1}{\ensuremath{#2}\xspace}}
  \newVar{\server}{S}
  \newVar{\client}{C}
  \newVar{\rclient}{R_c}
  \newVar{\rserver}{R_s}
  
\providecommand{\vs}{{vs.}\xspace}
\providecommand{\ie}{{i.e.,}\xspace}
\providecommand{\eg}{{e.g.,}\xspace}

\usepackage{algorithm}
\usepackage{algpseudocode}
\algtext*{EndWhile}
\algtext*{EndIf}
\algtext*{EndFor}
\algtext*{EndFunction}
\algtext*{EndProcedure}
\newcommand{\name}{\textsc{Spectra}\xspace}
\newcommand{\base}{\textsc{Baseline}\xspace}
\newcommand{\less}{\textsc{Less}\xspace}
\newcommand{\eclipse}{\textsc{Eclipse}\xspace}
\newcommand{\birkhoff}{\textsc{Birkhoff+}\xspace}
\newcommand{\wwfa}{\textsc{WWFA}\xspace}

\newcommand{\decomp}{\textsc{Decompose}\xspace}
\newcommand{\sched}{\textsc{Schedule}\xspace}
\newcommand{\equal}{\textsc{Equalize}\xspace}

\newcommand{\refine}{\textsc{Refine}\xspace}

\newcommand{\LB}{LB}
\newcommand{\D}{D}

\newcommand{\Drem}{D_{rem}}
\newcommand{\Srem}{S_{rem}}

\usepackage{orcidlink}

\ifcomm
	\newcommand{\mycomm}[3]{{\footnotesize{{\color{#2} \textbf{[#1: #3]}}}}}
     \newcommand{\Fmycomm}[3]{\footnote{{{\color{#2} \textbf{[#1: #3]}}} }}
     
\else
    \newcommand{\mycomm}[3]{}
    \newcommand{\Fmycomm}[3]{}
    
\fi

\newcommand{\I}[1]{}

\ifieee %
    \newcommand{\bp}{\begin{IEEEproof}}     %
    \newcommand{\bpo}{ \begin{IEEEproof}[Proof Outline] }
    \newcommand{\ep}{\end{IEEEproof}}       %
    \newcommand{\proofof}[1]{\begin{IEEEproof}[Proof of #1]} %
\else
    \newcommand{\bp}{\begin{proof}}
    \newcommand{\bpo}{ \begin{proof}[Proof Outline] }
    \newcommand{\ep}{\end{proof}}       %
    \newcommand{\proofof}[1]{\begin{proof}[Proof of #1]} %
\fi

\begin{document}

\title{Scheduling Parallel Optical Circuit Switches for AI Training
} 

\author{Kevin Liang, Litao Qiao, Isaac Keslassy\textsuperscript{\orcidlink{0000-0001-6103-6910}},~\IEEEmembership{Senior Member,~IEEE} and Bill Lin\textsuperscript{\orcidlink{0000-0003-0965-7247}},~\IEEEmembership{Senior Member,~IEEE}\thanks{K. Liang, L. Qiao and B. Lin are with UC San Diego (e-mails: \{k3liang, l1qiao, billlin\}@ucsd.edu). I. Keslassy is with the Technion and UC Berkeley (isaac@technion.ac.il). This work was partly supported by the Louis and Miriam Benjamin Chair in Computer-Communication Networks, NSF Award No.~2410053, and the Israel Science Foundation (grant No. 2895/26). A preliminary version of this work appeared in IEEE Networking Letters~\cite{liang-net}.
\ifarxiv
\protect\\ This work has been accepted for publication in the IEEE/Optica Journal of Lightwave Technology. %
\copyright~2026 IEEE. Personal use permitted; republication requires IEEE permission.
\fi}}
\ifarxiv\else
\markboth{JOURNAL OF LIGHTWAVE TECHNOLOGY}{Liang \MakeLowercase{\textit{et al.}}: Scheduling Parallel Optical Circuit Switches for AI Training}
\fi

\maketitle

\begin{abstract}
The rapid growth of AI training has dramatically increased datacenter traffic demand and energy consumption, which has motivated renewed interest in optical circuit switches (OCSes) as a high-bandwidth, energy-efficient alternative for AI fabrics. Deploying multiple parallel OCSes is a leading alternative. However, efficiently scheduling time-varying traffic matrices across parallel optical switches with non-negligible reconfiguration delays remains an open challenge.

We consider the problem of scheduling a single AI traffic demand matrix $D$ over $s$ parallel OCSes while minimizing the makespan under reconfiguration delay $\delta$. Our algorithm \name relies on a three-step approach: Decompose $D$ into a minimal set of weighted permutations; Schedule these permutations across parallel switches using load-aware assignment; then Equalize the imbalanced loads on the switches via controlled permutation splitting. Evaluated on realistic AI training workloads (GPT model and Qwen MoE expert routing) as well as standard benchmarks, \name vastly outperforms a baseline based on state-of-the-art algorithms, reducing schedule makespan by an average factor of $1.4\times$ on GPT AI workloads, $2.7\times$ on MoE AI workloads, and $2.4\times$ on standard benchmarks. Further, the makespans achieved by \name consistently approach newly derived lower bounds.
\end{abstract}


\section{Introduction}

\IEEEPARstart{D}{atacenter} networks have had trouble scaling at the speed of large-scale AI training compute, especially lagging the AI bandwidth and power constraints~\cite{liao2025mixnet,meta,Stellar}. 
Unlike traditional cloud workloads, AI training compute produces sustained high bisection bandwidth demand with repeated collective communication patterns across a large number of accelerators~\cite{alibaba}. Due to its tight iteration-level synchronization constraints, it is extremely sensitive to the worst-case completion time, \ie the collective makespan also known as collective completion time (CCT).
These properties stress networks based on conventional electronic packet-switched fabrics, whose power and bandwidth abilities scale poorly with network size~\cite{liao2025mixnet}.

There
has been a significant body of literature
proposing the use of optical circuit switches (OCSes) in datacenters as a compelling alternative~\cite{liao2025mixnet,feng2025railx,eclipse,solstice,
farrington2010helios,valls2021birkhoff}.
In comparison with electronic packet switches, OCSes provide much higher bandwidth at much lower per-bit energy, both of which are important in meeting the rapidly increasing demands on AI datacenter networks. 
To increase aggregated capacity, prior work increasingly considers employing \textit{parallel OCSes}~\cite{LESS,negotiator,zu2024resiliency,unlocking,vermilion}. On the other hand, OCSes suffer from reconfiguration delays that can significantly affect the CCT of the AI training collectives.

In this paper, we consider the problem of  scheduling an AI traffic demand matrix over a set of parallel OCSes with reconfiguration delays, while minimizing the makespan (and therefore the CCT) across the parallel switches.

\T{Related work.} 
The existing literature mostly minimizes schedule length either assuming a single optical switch, or a multi-stage switch architecture that can be {abstracted as a single switch}~\cite{eclipse,solstice,farrington2010helios,valls2021birkhoff}.
In particular, these existing works typically focus on the scheduling of a given traffic demand matrix $D$ onto an OCS by decomposing $D$ into a sequence of configurations of given durations, 
such that the traffic demands from $D$ are served.
A critical difference with an analogous decomposition problem for electronic switches is that OCSes incur a nontrivial reconfiguration delay $\delta$ when the \emph{switch configuration} has to change, which adds to the total delay (the \emph{makespan}) in servicing $D$.
Existing works (\eg \cite{eclipse,solstice,farrington2010helios}) are effective in solving this scheduling problem with reconfiguration delays for the single OCS case.

\less~\cite{LESS} is the most relevant work on scheduling a demand matrix $D$ for $s$ parallel OCSes. It splits 
$D$ into $s$ sub-matrices 
using a balancing criterion  with a sparsity objective, then schedules each sub-matrix on a separate OCS. {However, \less focuses more on a partial-reconfiguration scheduling algorithm intended for partially-reconfigurable switches.}
Additional works consider parallel OCSes, but under different assumptions: They (1)~achieve a distributed scheduling~\cite{negotiator}, but with lower matching efficiency; (2)~use OCSes to route around failures~\cite{zu2024resiliency}; (3)~extend results to a multi-hop schedule in which nodes can serve as intermediate nodes for other flows~\cite{unlocking,sirius,topoopt}, which we do not consider in this paper; and (4)~focus on integer-weight decompositions~\cite{vermilion}.

\T{Contributions.} 
We introduce the \name
algorithm for scheduling an AI demand matrix $D$ across $s$ parallel switches with reconfiguration delay, while minimizing a CCT-based makespan objective. 
A key idea of \name is to cut the problem into three successive sub-problems, each practically solved with polynomial-time algorithms:
(i)~\decomp $D$ into a set of permutations, (ii)~\sched these permutations across the $s$ switches via load-aware scheduling, then (iii)~\equal the loads on the switches by splitting the durations of some of the permutations on the most-loaded switches and moving part of their load to the least-loaded ones. 
We 
rigorously derive new lower bounds on the achievable makespan for arbitrary demand matrices by any parallel OCS scheduling algorithm.
We also introduce a new Qwen-57B Mixture of Experts (MoE) workload by measuring its traffic in a 64-GPU cluster. We then compare \name against a \base algorithm that extends the sparsity ideas of \less~\cite{LESS}, as well as against an \eclipse~\cite{eclipse}-based version of \name, using three workloads: (i)~a GPT-based AI training model, (ii)~the new Qwen MoE measurements, and (iii)~a standard benchmark in the literature.
\name
vastly outperforms \base, with a schedule that is on average shorter by a factor of $1.4\times$ on GPT AI workloads, $2.7\times$ on MoE AI workloads, and $2.4\times$ on standard benchmarks. It also always outperforms the \eclipse-based variant.
More importantly, the makespans achieved by \name are consistently approaching the theoretically derived lower bounds, thus achieving near-optimal performance in practice with fast polynomial-time algorithms, with robust performance across sparse and dense traffic regimes.
Our results show that algorithmic co-design of
traffic decomposition and parallel OCS scheduling is critical
for improving optical circuit switching for next-generation AI
datacenters.
Our code is available at \url{https://github.com/k3liang/SPECTRA}.

\begin{figure}
    \centering
    \includegraphics[width=0.8\linewidth]{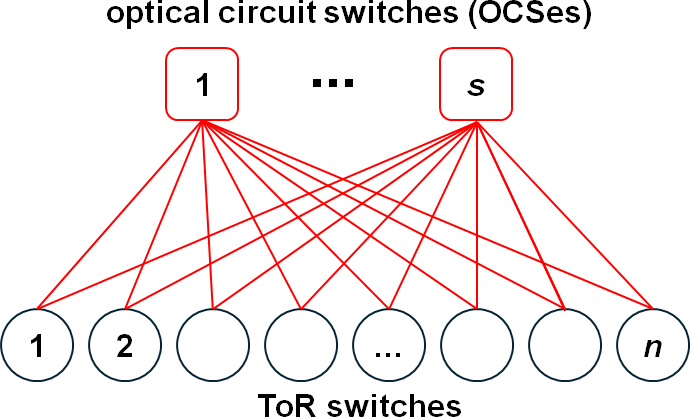}
  \caption{Datacenter network topology with $s$ parallel OCSes.}
  \label{fig:topology}
\end{figure}

\section{Problem Formulation}
\label{sec:formulation}

\T{Demand matrix.} As \cref{fig:topology} illustrates, we consider a datacenter core of $s$ parallel \textit{optical circuit switches} (OCSes). Each OCS has $n$ input and output ports, connected to the Top-of-Rack (ToR) switches of $n$ datacenter racks. 
Periodically, a centralized controller computes a new $n \times n$ traffic demand matrix $D$. Each $D_{ij}$ corresponds to the  normalized volume of data that needs to be switched from rack $i$ to rack $j$ over the next period.
We assume that $D$ is doubly substochastic, \ie each row and column sums up to at most one. We call $D$ doubly stochastic when they all sum up to exactly one.

\T{Switch schedule.} Each OCS has a schedule that consists of a sequence of permutations, where each input rack $i$ is connected to a single output rack $j$. A sequence of $k$ permutations $P_1, P_2, \ldots, P_k$, with corresponding weights $\alpha_1, \alpha_2, \ldots, \alpha_k > 0$ 
is said to \emph{cover} $D$ if the following is satisfied:
\begin{equation} \label{eq:Dcover}
\sum_{i=1}^{k} \alpha_i P_i \geq D.
\end{equation}
In addition, the optical switch needs a \textit{reconfiguration delay} $\delta$ before each permutation. Thus, the time needed to schedule the $k$ permutations in a single switch is 
\begin{equation} \label{eq:schedule-delay}
\sum_{i=1}^{k} (\delta + \alpha_i) = k \cdot \delta + \sum_{i=1}^{k} \alpha_i  .
\end{equation}

\T{Makespan.} We now consider the $s$ schedules of the $s$ optical switches. As above, consider some switch $h$, with $h \in \sbrac{1,s}$. Assume that each such switch uses a schedule with $k^h$ 
permutations $P_i^h$ and their corresponding weights $\alpha_i^h$, for $i \in \sbrac{1,k^h}$. Then, extending~\cref{eq:Dcover}, the set of all $s$ switch schedules covers $D$ {when}
\begin{equation}
\label{eq:parcover}
\sum_{h=1}^s \sbrac{ 
    \sum_{i=1}^{k^h} \alpha_i^h P_i^h  } 
\geq D.
\end{equation}
The \emph{makespan} of the $s$ parallel switch schedules is the delay of the longest one, \ie using \cref{eq:schedule-delay},
\begin{equation}
\label{eq:parmakespan}
\max_{1 \leq h \leq s} \sbrac{
    \sum_{i=1}^{k^h} \para{
        \delta+\alpha_i^h } }.
\end{equation}

\T{Objective.} The goal of this paper is to find all permutations and weights in the $s$ schedules of the $s$ parallel optical switches, 
so that they manage to \textit{cover $D$}  (\cref{eq:parcover}) while \textit{minimizing the makespan}.

The general minimization problem of the above objective is NP-hard, even with $s=1$~\cite{hamdi}. Thus, in this paper, we focus on finding a practical algorithm that typically \textit{attains a small makespan in practice}.

\section{Algorithms}
\label{sec:algorithms}

\subsection{Overview}

In this section, we introduce our \name algorithm (short for \textbf{S}cheduling \textbf{p}arall\textbf{e}l \textbf{c}ircuit switches for data center \textbf{tra}ffic). 
As formally defined in \cref{sec:formulation}, \name finds $s$ schedules of $D$ over $s$ parallel optical switches, such that the schedules \emph{cover} $D$ while attaining a  \emph{small makespan}. We start by illustrating \name's three-step approach using an example, before formally defining each step.

\T{\decomp.}
The first step in \name is to \decomp a demand matrix $D$ into a set of permutations $\{P_1, P_2, \ldots, P_k\}$ with a corresponding set of weights $\{\alpha_1, \alpha_2, \ldots, \alpha_k\}$ such that their weighted sum \emph{covers} $D$, \ie $\sum_{i=1}^k \alpha_i P_i \geq D$.

\begin{figure}
    \centering
    \includegraphics[width=0.5\linewidth]{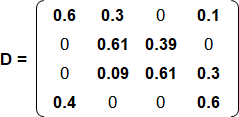}
    \caption{Example of demand matrix $D$.}
    \label{fig:toy}
\end{figure}

\begin{figure*}
    \centering
    \includegraphics[width=\linewidth]{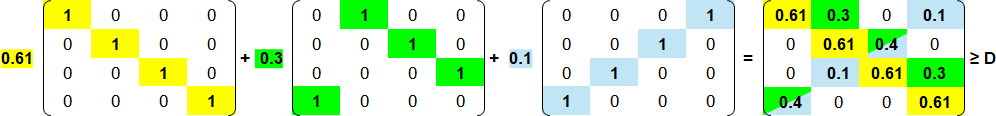}
    \caption{Example decomposition into $k = 3$ weighted permutations that cover $D$.}
    \label{fig:toy-decomp}
\SmallReduceVSpace \end{figure*}

Consider the demand $D$ shown in \Cref{fig:toy}. \Cref{fig:toy-decomp} provides an example of how $D$ can be decomposed into 3 permutations, with the corresponding weights 0.61, 0.3, and 0.1. \Cref{fig:toy-decomp} also shows that their weighted sum indeed covers $D$. As detailed later in this section, given at most $k$ nonzero elements in any row or column of $D$, our \decomp algorithm \textit{guarantees} that exactly $k$ permutations will be generated to cover $D$. It minimizes the number of reconfiguration delays needed while also reducing the sum duration of the permutations.

\T{\sched.}
After the \decomp step, we next \sched the $k$ permutations across the $s$ switches. This step is equivalent to scheduling $k$ jobs across $s$ machines, with each $\alpha_i$ representing the amount of time that a job (permutation) $P_i$ needs to run, with added consideration for a reconfiguration delay $\delta$ each time a machine (switch) needs to be (re)configured to run a job (switch permutation).

\begin{figure*}
    \centering
    \includegraphics[width=\linewidth]{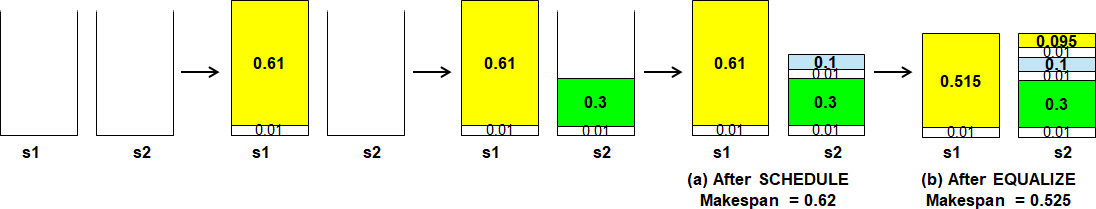}
    \caption{(a) Scheduling the $k = 3$ permutations across $s = 2$ switches. (b) Equalizing the loads among the switches. }
    \label{fig:schedule}
\SmallReduceVSpace \end{figure*}

For this step, 
we first sort the permutations by non-increasing $\alpha_i$ weights, and then we schedule the corresponding $P_i$ in non-increasing-weight-order to the \emph{least} loaded switch each time. 
\Cref{fig:schedule} illustrates the \sched process for the 3 permutations from \Cref{fig:toy-decomp} across $s = 2$ switches with reconfiguration delay $\delta = 0.01$. Initially, both switches are empty, as shown in \Cref{fig:schedule}. We first schedule the permutation with $\alpha_1 = 0.61$ to the first switch, so the load of the first switch becomes $0.61 + 0.01 = 0.62$. We next schedule the permutation with $\alpha_2 = 0.3$ to the second switch since it is empty and hence the least loaded. The load of the second switch becomes $0.3 + 0.01 = 0.31$. Finally, we schedule the last permutation with $\alpha_3 = 0.1$ to the second switch again, since it remains the least loaded of the two switches, making its final load $0.42$. The makespan after the \sched step is $\max(0.62, 0.42) = 0.62$, as shown in \Cref{fig:schedule}(a).

\T{\equal.}
As the final step, we aim to \equal the loads on the switches by splitting the duration ($\alpha_i$) of a permutation (switch configuration) $P_i$ into \emph{multiple} segments and moving some segments to different switches in order to load-balance the workloads. In particular, our \equal step iteratively balances the loads between the \emph{most} loaded switch $h_{max}$ and the \emph{least} loaded switch $h_{min}$ by moving a portion of the longest-duration permutation in $h_{max}$ to $h_{min}$. This is shown in Fig.~\ref{fig:schedule}(b), where we split $\alpha_1 = 0.61$ in the first switch into two parts, an $\alpha_1^1 = 0.515$ part that remains in the first switch and an $\alpha_1^2 = 0.095$ part that moves to the second switch, resulting in the final makespan of $0.525$, as shown in \Cref{fig:schedule}(b).

Each of the above steps 
is further detailed below.

\subsection{\decomp}
\label{sec:decomp}

\T{Support matrix.}
Given an $n \times n$ matrix $D$, let $S\in \{0,1\}^{n \times n}$ be its \emph{support} matrix defined by $S_{ij} = 1$ if $D_{ij} > 0$ and $S_{ij} = 0$ otherwise.
Let $k$ be the \emph{degree} of $D$, corresponding to the maximum number of nonzero elements in any row or column in $D$.
By definition, the support matrix $S$ for $D$ will also have a degree of $k$, corresponding to a maximum of $k$ 1's in any row or column in $S$. The following property follows from König's Line Coloring Theorem: \begin{property}
\label{prop:k}
$k$ permutations are necessary and sufficient to cover any matrix of degree $k$.
\end{property}

\T{\decomp intuition.} As \cref{fig:schedule}(a) illustrates, the sum of all the schedule delays will include (1)~$\delta$ times \textit{the number of permutations} (in this example, $0.01 \cdot 3$), and (2)~\textit{the sum of the permutation weights} ($0.61+0.3+0.1=1.01$). Thus, we want to minimize both of these components. 

(1)~First, by Property~\ref{prop:k}, 
$k$ permutations can cover $D$ via some $\alpha_1, \alpha_2, \ldots, \alpha_k$ such that $\sum_{i=1}^k \alpha_i P_i \geq D$.
In our approach, the goal of \textit{minimizing the number of permutations} is to later minimize the number of reconfiguration delays.
Therefore, our approach is to decompose $D$ into exactly this minimum number $k$ of permutations.

(2)~However, there may be many possible sets of $k$ permutations that can cover $D$.
Among these, we want to find a set of permutations along with a set of parameters $\{\alpha_1, \alpha_2, \ldots, \alpha_k\}$ that also \textit{minimizes the total duration} $\sum_{i=1}^k \alpha_i$.

The goal of these two properties is to avoid a decomposition with many coefficients $\alpha_i$ that are smaller than  the reconfiguration delay $\delta$. The usual Birkhoff  decompositions~\cite{valls2021birkhoff} only target $\sum_{i}\alpha_i \leq 1$. However, this comes at the cost of the number of reconfigurations increasing significantly, with many small coefficients $\alpha_i$, thus also increasing the total makespan due to the additional reconfiguration delay $\delta$ for each configuration.
By bounding the number of reconfigurations to be exactly $k$, we may increase $\sum_i \alpha_i$ to slightly exceed one, but in return, reconfiguration cost and the makespan greatly decrease.

\begin{algorithm}[t]
\caption{{\decomp}}\label{alg:decomp}
\begin{algorithmic}[1]
\State \textbf{Input:} $D$ and $\delta$
\State \textbf{Assign:} $S = \text{support matrix of } D$, $\Drem = D$, $\Srem = S$, $i = 1$, $\mathcal{A} = \{ \}$ and $\mathcal{P} = \{ \}$ 
\While{$\Srem \neq \emptyset$}
\State $P_i \gets \text{MWM-Node-Coverage}(\Drem, \Srem)$
    \State $\alpha_i \gets \min_{(a, b), (P_i)_{ab} = 1} (\Drem)_{ab}$
    \State $\mathcal{P} \gets  \mathcal{P} \cup \{P_i\}$
    \State $\mathcal{A} \gets \mathcal{A} \cup  \{\alpha_i\}$
    \State $\Drem \gets \Drem - (\alpha_i P_i)$
    \State $\Srem \gets \Srem - (P_i \cap \Srem)$
    \State $i \gets i + 1$
\EndWhile
\State $\hat{\mathcal{A}} \gets \refine(D, \mathcal{A}, \mathcal{P})$
\State \textbf{return} $(\hat{\mathcal{A}}, \mathcal{P})$
\end{algorithmic}
\end{algorithm}

\T{\decomp pseudocode.}
\Cref{alg:decomp} shows the pseudocode of our \decomp algorithm.
Given $D$, we first derive its support matrix $S$.
\decomp then iteratively generates a set of $k$ permutations $\mathcal{P} = \{P_1, P_2, \ldots, P_k\}$ that cover $S$ by solving a \emph{maximum weight matching} (MWM) problem \emph{under node coverage constraints}~\cite{schrijver2003combinatorial}
in each iteration.

At a high-level, we track the \emph{remaining traffic} in $D$ and the \emph{remaining uncovered elements} in $S$ after each round, denoted as $\Drem$ and $\Srem$, respectively. They are initially $\Drem = D$ and $\Srem = S$. Based on the updated $\Drem$ and $\Srem$ in each round, we would like to generate a permutation that satisfies two properties. Based on the above intuition, (1)~we want to choose a permutation that is \emph{guaranteed} to reduce the degree of $\Srem$ in each round. This way, the algorithm will end after exactly $k$ rounds, generating exactly $k$ permutations, where $k$ is the degree of the initial support matrix $S$.
(2)~Second, among the permutations that will reduce the degree of $\Srem$ by 1, we want to choose a permutation that will reduce the amount of remaining traffic in $\Drem$ as much as possible.

Fortunately, this is a well-known optimization problem called \emph{MWM 
under node coverage constraints}~\cite{ahuja1993network,schrijver2003combinatorial}, 
where efficient polynomial-time algorithms already exist to satisfy our two desired properties.
In this formulation, we identify \emph{critical} rows and columns in $\Srem$, which are the rows and columns that have $k$ nonzero entries, where $k$ is the degree of $\Srem$.
To ensure that the degree of $\Srem$ gets reduced by 1, we \emph{constrain the matching of every critical row or column to a nonzero entry in that row or column}.
Among the permutations that satisfy this constrained matching, we use $\Drem$ as weights to optimize the MWM.
This constrained MWM problem can be solved 
in polynomial time using the Hungarian algorithm~\cite{schrijver2003combinatorial}, modified to handle the above coverage constraints.
This is shown in Line 4 of~\cref{alg:decomp}.

For the generated $P_i$, we derive the initial weight $\alpha_i$ by taking the minimum entry $\para{\Drem}_{ab}$ where the corresponding entry $\para{P_i}_{ab}$ is a 1 (Line 5).
We then update $\Drem$ and $\Srem$ at the end of each round accordingly (Lines 8-9), and we repeat the while-loop until all 1-elements in $S$ are covered by a permutation in $\mathcal{P}$. Note that $\Srem$ at each round corresponds to the remaining \textit{uncovered} elements of $S$ till that round, and not necessarily to the support of $\Drem$.

\begin{algorithm}[t]
\caption{Subroutine \refine}\label{alg:refine}
\begin{algorithmic}[1]
\State \textbf{Input:} $D$, $\mathcal{A}$ and $\mathcal{P}$
\State $\hat{\mathcal{A}} \gets \mathcal{A}$
\State $\Drem \gets D - \left( \sum_{i=1}^k \alpha_i P_i \right)$
\For{$i = 1 \text{ to } k$}
    \State $d \gets \max_{(a, b), (P_i)_{ab} = 1} (\Drem)_{ab}$
    \State $\hat{\alpha}_i \gets \hat{\alpha}_i + d$
\For{all $(a, b) \text{ such that } (P_i)_{ab} = 1$}
        \State $(\Drem)_{ab} \gets \max(0, (\Drem)_{ab} - d)$
    \EndFor
\EndFor
\State \textbf{return} $\hat{\mathcal{A}}$
\end{algorithmic}
\end{algorithm}

\T{\refine subroutine.}
After the while-loop in Lines 3-10 of~\cref{alg:decomp}, we are guaranteed that the set of permutations $\mathcal{P} = \{P_1, P_2, \ldots, P_k\}$ will cover $S$, but the weighted sum with the corresponding $\mathcal{A} = \{\alpha_1, \alpha_2, \ldots, \alpha_k\}$ may not necessarily cover $D$. This is because in Line 5, we are taking the \emph{minimum} value in $\Drem$ covered by $P_i$, but our stopping criterion is just that the initial support matrix $S$ is covered by the set of permutations.
The role of the final \refine step in Line 11 of~\cref{alg:decomp} is to refine the set of weights to ensure that the weighted sum indeed covers $D$.

Given a set of $k$ permutations $\mathcal{P} = \{P_1, P_2, \ldots, P_k\}$ that covers $S$, we could readily derive a set of new weights $\hat{\mathcal{A}} = \{\hat{\alpha}_1, \hat{\alpha}_2, \ldots, \hat{\alpha}_k\}$ 
such that the weighted sum covers $D$, by solving the following linear programming (LP) problem:
\be
\label{eq:LP2}
    \min \sum_{i=1}^k \hat{\alpha}_i, \quad\quad\quad 
    \text{s.t. } \sum_{i=1}^k \hat{\alpha}_i P_i \geq D .
\ee

Instead, starting from the initial weights $\mathcal{A} = \{\alpha_1, \alpha_2, \ldots, \alpha_k\}$ from Lines 3-10, we greedily increase the weights to a new set of weights $\hat{\mathcal{A}}$ such that the weighted sum covers $D$ (see \Cref{alg:refine}).
We find that in practice this greedy approach works just as well as solving the LP.

\subsection{\sched}

After the \decomp step, we have a set of permutations $\mathcal{P}$ 
and their corresponding weights 
$\hat{\mathcal{A}}$. 
In the \sched step, we want to assign each of these permutations to one of the $s$ switches to minimize the makespan (\cref{eq:parmakespan}). This is equivalent to the makespan minimization problem on identical parallel machines, for which the \textit{Longest Processing Time (LPT) First} algorithm~\cite{berge2024approximation} 
is a well-known and effective greedy heuristic. We adopt it for our \sched step (\cref{alg:sched}). 

\begin{algorithm}[t]
\caption{{\sched}}\label{alg:sched}
\begin{algorithmic}[1]
\State \textbf{Input:} $\mathcal{A}$, $\mathcal{P}$, $s$ and $\delta$
\State Sort $\mathcal{P}$ by non-increasing weights in $\mathcal{A}$ such that $\alpha_{\sigma(1)} \geq \alpha_{\sigma(2)} \geq \cdots \geq \alpha_{\sigma(k)}$
\For{$h = 1 \text{ to } s$}
    \State $L_h \gets 0$, $\mathcal{P}^h \gets \{ \}$, $\mathcal{A}^h \gets \{ \}$
\EndFor
\For{$i = 1 \text{ to } k$}
    \State $h^* = \argmin_h L_h$
    \State $\mathcal{P}^{h^*} \gets \mathcal{P}^{h^*} \cup P_{\sigma(i)}$
    \State $\mathcal{A}^{h^*} \gets \mathcal{A}^{h^*} \cup \alpha_{\sigma(i)}$ 
    \State $L_{h^*} \gets L_{h^*} + \delta + \alpha_{\sigma(i)}$
\EndFor
\State $\mathcal{S} \gets \{ (\mathcal{A}^1, \mathcal{P}^1), 
(\mathcal{A}^2, \mathcal{P}^2),
\ldots, (\mathcal{A}^s, \mathcal{P}^s)\}$
\State \textbf{return} $\mathcal{S}$
\end{algorithmic}
\end{algorithm}

\T{\sched pseudocode.}
In this greedy algorithm, we first sort the permutations by non-increasing weights, as shown in Line 2. We then initialize the load $L_h$ for all switches to 0 with no permutations assigned to switch $h$ yet, as shown in Lines 3-4. Then in Lines 5-9, following a non-increasing order of weights, we assign in a greedy manner the corresponding permutation to the least loaded switch (Line 6), add the permutation to the set of permutations already assigned to that switch (Lines 7-8), and update the load of that switch, including the reconfiguration delay $\delta$ (Line 9). In the end, the procedure returns the schedule $\mathcal{S}$ of the $s$ switches in the form of weights and permutations $(\mathcal{A}^h, \mathcal{P}^h)$ per switch.

\subsection{\equal}

Finally, \cref{fig:schedule}(b) illustrates how
we can reduce the makespan by redistributing some of the switching load from one switch to another. 

\begin{algorithm}[t]
\caption{{\equal}}\label{alg:equal}
\begin{algorithmic}[1]
\State \textbf{Input:} $\mathcal{S} = \{ (\mathcal{A}^1, \mathcal{P}^1), 
(\mathcal{A}^2, \mathcal{P}^2),
\ldots, (\mathcal{A}^s, \mathcal{P}^s)\}$, $s$ and $\delta$
\For{$h = 1 \text{ to } s$}
    \State $L_h \gets \sum_{i=1}^{k^h} (\delta + \alpha_i^h)$
\EndFor
\State cont $\gets$ \textsc{true}
\State num\_iterations $\gets$ $0$
\While{cont \textbf{and} num\_iterations $< ks$}
    \State num\_iterations $\gets$ num\_iterations $+ ~1$
    \State $h_{max} = \argmax_h L_h$
    \State $h_{min} = \argmin_h L_h$
    \If{$\left( L_{h_{max}} - L_{h_{min}} \right) > \delta$}
        \State $\mu \gets \left( L_{h_{max}} + L_{h_{min}} + \delta \right) / 2$
        \State $z \gets \argmax_i \alpha_i^{h_{max}}$
        \If{$\alpha_z^{h_{max}} > \left( L_{h_{max}} - \mu \right)$}
            \State $\tau \gets L_{h_{max}} - \mu$
            \State $\alpha_z^{h_{max}} \gets \alpha_z^{h_{max}} - \tau$
            \State Copy permutation $P_z^{h_{max}}$ to $\mathcal{P}^{h_{min}}$ 
            \State Add $\tau$ as corresponding weight to $\mathcal{A}^{h_{min}}$
            \State $L_{h_{max}} \gets L_{h_{max}} - \tau$
            \State $L_{h_{min}} \gets L_{h_{min}} + \delta + \tau$
        \Else
            \State cont $\gets$ \textsc{false}
        \EndIf
    \Else
        \State cont $\gets$ \textsc{false}
    \EndIf
\EndWhile
\State $\hat{\mathcal{S}} \gets \text{updated } \{ (\mathcal{A}^1, \mathcal{P}^1), 
(\mathcal{A}^2, \mathcal{P}^2),
\ldots, (\mathcal{A}^s, \mathcal{P}^s)\}$
\State \textbf{return} $\hat{\mathcal{S}}$
\end{algorithmic}
\end{algorithm}

\T{\equal pseudocode.}
The pseudocode for this \equal step is shown in~\cref{alg:equal}.
We first compute the loads $L_h$ for each switch after the \sched step in Lines 2-3.
We then iteratively equalize the loads in Lines 6-23 by moving a portion of the load from the most loaded switch $h_{max}$ to the least loaded switch $h_{min}$. Lines 8-9 identify $h_{max}$ and $h_{min}$. If the difference between them is more than $\delta$, we compute the target load $\mu$ to be the average of the two loads that includes a $\delta$ delay when a portion of $h_{max}$ is moved to $h_{min}$ (Lines 10-11).
In the most loaded switch $h_{max}$, we identify the index $z$ of the longest-duration permutation in $h_{max}$ in Line 12. If the weight of this permutation $\alpha_z^{h_{max}}$ is more than the difference between the current and target load of $h_{max}$, then we move the difference $\tau$ from $\alpha_z^{h_{max}}$ in $h_{max}$ to $h_{min}$ by copying the corresponding permutation $P_z^{h_{max}}$ to switch $h_{min}$ and assigning a weight of $\tau$ to it (Lines 13-17). We then adjust the loads of the two switches accordingly in Lines 18-19.
The iterations continue as long as load balancing is possible.

\subsection{Time Complexity}

\T{\decomp} relies on a loop with $k$ iterations, based on the support matrix logic. At each iteration, the MWM-Node-Coverage takes $O(n^{3})$ steps~\cite{gabow2018data}, 
yielding $O(kn^{3})$ overall. 
In practice, the performance can be sped up on average cases through a Jonker-Volgenant implementation~\cite{jonker1999linear, jonker}. The \refine subroutine goes through $k$ iterations, each corresponding to a particular weight $\alpha_i$. Each iteration requires finding a maximum of up to $n$ entries in $D_{rem}$, and up to $n$ updates to $D_{rem}$, which results in $O(kn)$ complexity. Therefore, the total time complexity of \decomp is bounded by $O(k n^3)$.

\T{\sched} requires an $O(k\log k)$ sorting of $\mathcal{A}$. There are also $k$ main iterations, with each iteration requiring an $\argmin_h L_h$. If we store the group of $L_h$ as a heap of size $s$, then an iteration will be $O(\log s)$. This yields a total of $O(k\log k + k\log s)=O(k\log (ks))$.

\T{\equal} has $\argmax_h L_h$, $\argmin_h L_h$, and $\argmax_i \alpha_i^{h_{max}}$ in the loop. If we keep heaps to store these collections, then each loop iteration would take $O(\log k + \log s)$ steps. The total number of iterations of the loop is bounded by $ks$, thus
the total time complexity is $O(ks\log (ks))$. In practice, \equal uses significantly fewer than $ks$ iterations.

\section{Lower Bounds} 
\label{sec:LBs}

In this section, we provide new  
lower bounds for the scheduling makespan of any
demand matrix $D$ using $s$  optical switches and a reconfiguration cost of $\delta >0$. 

\T{Notations.} We index the rows and columns of $n\times n$ matrix $D$ using $i\in \sbrac{1,2n}$, such that the first $n$ indices cover the $n$ rows and the next $n$ indices correspond to the $n$ columns. 

Given some row or column of $D$, we provide below several lower bounds using different approaches. Then, the maximum of all these lower bounds over all rows/columns is also a lower bound for the makespan. %

\begin{property}
Assume that an alternative demand matrix $D_i$ constructed with the same row/column $i\in \sbrac{1,2n}$ as $D$ and zero everywhere else yields $L$  lower bounds $\LB_i^{(l)}$ for $l\in \sbrac{1,L}$ on its schedule makespan. Then a lower bound on the schedule makespan for $D$
is 
\be
LB = \max_{i\in \sbrac{1,2n}} \max_{l\in \sbrac{1,L}} {\LB_i^{(l)}}.
\ee
\end{property}
\bp
To schedule $D$ we need to schedule at least its row/column $i$, so $\LB_i^{(l)}$ is a lower bound on the makespan for $D$. Moreover, the finite cartesian product of finite sets also has a finite cardinality; and for any finite set of finite numbers, the maximum element exists, is within that set, and is also finite. Thus, the maximum of the lower bounds is a lower bound itself, and it will be one of these lower bounds $\LB_i^{(l)}$.
\ep

\T{Lower bound 1.} We start by demonstrating a general lower bound for any number of elements per row/column.

\begin{theorem}[Lower bound 1]
\label{thm:lb1}
Assume that row/column $i$ has $k_i$
nonzero elements and a total weight $w_i$. Then the scheduling makespan has lower bound 
\be
\LB_i^{(1)} = \frac{w_i+\delta \cdot \max \para{k_i,s}}{s} 
\ee
\end{theorem}
\bp
Any set of $s$ schedules that we produce needs to schedule at least the $k_i$ nonzero elements of row/column $i\in \set{1,\dots,2n}$. To determine a lower bound, we assume that this is all we need to service and neglect the other elements in the demand matrix. We need at least $k_i$ scheduling configurations to schedule $k_i$ elements. Their total weight also needs to be at least $w_i$. Thus, 
the sum of the $s$ schedule lengths will be at least $w_i$ plus the reconfiguration times, \ie 
$w_i+k_i \cdot \delta$.
Since we have $s$ parallel schedules, the worst-case schedule duration is at least the 
average schedule duration, 
\ie at least $\frac{w_i+k_i \cdot \delta}{s}$. %

In addition, %
we need to schedule $w_i$ over the $s$ switches, yielding a $w_i/s$ average delay, and each switch needs at least $\delta$ reconfiguration time. Thus another lower bound is $\frac{w_i}{s}+\delta$. Combining both, we get the result.
\ep

\T{Example.} Assume that in row/column $i$ of doubly stochastic  demand matrix $D$, there are $k_i=16$ nonzero elements. If $s=4$, then our lower bound is 
$\LB_i^{(1)} = \frac{1+16\cdot \delta}{4}= 1/4 + 4 \cdot \delta.$

\T{Lower bound 2.} We now prove a second lower bound for the special case where the number $k_i$ of nonzero elements in row or column $i$ exactly equals $s$. We start with a useful lemma.

\begin{lemma}
\label{lem:1}
If row/column $i$ has $k_i=s$ nonzero elements, and there is a total of $s+m$ configurations in the $s$ parallel schedules, then the $s$ parallel schedules will necessarily service at most $m$ element chunks. 
In other words, at least $s-m$ whole elements will be serviced within a single configuration and will not be spread between configurations.
\end{lemma}
\bp
We will prove the lemma by induction on $m$. First, %
if there are no reconfigurations at all ($m=0$) and the $s$ nonzero elements are scheduled by the $s$ configurations of the $s$ parallel schedules, then necessarily each element is assigned to a single distinct configuration and single distinct schedule. That is, they cannot be combined within any schedule in any way. Otherwise, if one of them were {cut} into two chunks that are serviced by different configurations and/or schedules, then we would have $s+1$ chunks serviced within the $s$ schedules. Thus, by the pigeonhole principle, at least one schedule would service at least two chunks, which would need a reconfiguration. Contradiction.

Assume that the statement holds for $m\geq 0$, and let's prove it for $m+1 \geq 1$ reconfigurations. (1)~Assume that the last reconfiguration of at least one schedule out of the $s$ schedules is followed by (a)~an element chunk or (b)~a zeroed element. Then by removing the reconfiguration and (a)~appending the element chunk to another chunk of this element, or (b)~throwing out the zeroed element, we %
get a valid schedule with $m$ reconfigurations. By the inductive hypothesis, %
this alternative valid schedule would have at most $m$ chunks, and therefore we can conclude that our real schedule has at most $m+1$ chunks. (2)~Else, if only full nonzero elements follow reconfigurations, either (a)~all nonzero elements are full and there are no element chunks, which proves the result, or (b)~at least one schedule starts with a nonzero element chunk. Permute this nonzero element chunk with any nonzero full element that follows a reconfiguration, and we get back to case (1). Thus all cases hold
for $m+1$.
\ep

\begin{theorem}[Lower bound 2]
\label{thm:lb2}
Assume that row/column $i$ has $k_i=s$
nonzero elements $x_1 \geq x_2 \geq \cdots \geq x_{s}$, and a total weight $w_i$. 
Then the scheduling makespan has lower bound 
{\small
\begin{align} 
\LB_i^{(2)} = \delta+\min \Big(
& x_1,  \max \para{x_2,\frac{w_i+1\cdot\delta}{s}, x_{s}+\delta}, \nonumber\\
& \min_{2\leq m \leq s^2} \max \para{x_{m+1},\frac{w_i+m\cdot\delta}{s}}  \Big).
\end{align}
}
\end{theorem}
\bp %
To establish the lower bound, we assume again that we only want to schedule the nonzero elements of row/column $i\in \set{1,\dots,2n}$, and neglect all the other elements in $D$. %
We can use \cref{lem:1}.
With no reconfigurations, all elements are serviced as whole elements (within a single configuration) and do not need to be split into several chunks. %
Thus the makespan is at least the weight of the strongest element $x_1$, since it cannot be {split}, together with its configuration cost $\delta$. Likewise, with $m$ reconfigurations, at least $s-m$ elements will be serviced within a single configuration, and only $m$ elements can be split. In particular, at least one element of weight at least $x_{m+1}$ will be serviced within a single configuration, so the makespan is at least $x_{m+1}+\delta$.

Assume that our optimal schedule uses $m$ reconfigurations. The theorem successively considers all possible numbers $m$ of reconfigurations, and the makespan beyond the initial configuration cost of $\delta$: \\
\noindent (0)~Assume we have no reconfiguration at all. Then we saw that the makespan is at least $\delta+x_1$. Note that adding the average weight $w_i/s$ to $\delta$ would also be a lower bound for the makespan (as also shown in \cref{thm:lb1} using $k_i=s$), but since $x_1$ is the maximum weight, $x_1 \geq w_i/s$ so it forms a tighter lower bound.\\
\noindent (1)~Assume we have exactly $m=1$ reconfiguration. Then after the initial configuration cost of $\delta$, the makespan is at least (a)~$x_2$, as we saw; 
(b)~the average schedule length, given that the total schedule length is the sum of the weight $w_i$ and the reconfiguration delay $\delta$, and that there are $s$ schedules; and
(c)~the weight $x_s$ of the smallest element together with the reconfiguration delay $\delta$. This is because we saw that at most one element is {cut}. Consider the unique schedule with the reconfiguration delay. If both of its configurations are two chunks of the same cut element, then we can just get a better schedule by merging them and removing the reconfiguration, so this cannot happen. So at least one of the two elements in the schedule with the reconfiguration delay is not {cut}, thus its weight is at least $x_s$. To conclude with $m=1$, since the makespan is at least all of these three options, it is also at least their maximum. \\
\noindent ($m$)~Assume we have $m \geq 2$ reconfigurations. Then after the initial %
$\delta$, the makespan is at least (a)~$x_{m+1}$, as we saw; and 
(b)~the average schedule length, given that the total schedule length is the sum of the weight $w_i$ and the $m$ reconfiguration delays of $\delta$ each, and that there are $s$ schedules. Thus it is at least the maximum of these two options.

Finally, since two chunks of the same element within the same schedule can always be merged, the optimal number of reconfigurations is no more than $s^2$, and therefore bounded. Moreover, an optimal number of reconfigurations with a minimal makespan could be any $m$, so the minimum of these values forms a lower bound. 
\ep

Note that when its condition $k_i=s$ holds, the lower bound from \cref{thm:lb2} is at least as good as the first one from \cref{thm:lb1}: $\LB_i^{(2)} \geq \LB_i^{(1)}$. Furthermore, it is strictly better when not all nonzero elements are equal. 
Specifically, the first lower bound becomes $\LB_i^{(1)} = \frac{w_i}{s}+\delta$. 
The proof details the case with $m=0$ reconfigurations, 
and in the other cases the inequality is strict as $\delta+\frac{w_i+m\cdot\delta}{s} > \delta+\frac{w_i}{s}$ for $m>0$. 

\T{Demand matrix as a sum of weighted permutations.} We are interested in a special case where the $n \times n$ demand matrix $\D$ is a sum of (weighted) permutations. This case appears for example in two workloads: (1) the benchmark workload used in the evaluations; and (2) a Fully Sharded Data Parallel (FSDP2)~\cite{rajbhandari2020zero,zhao2023pytorch} workload with HSDP configuration using random server placement: FSDP2 traffic is implemented as a hierarchical ring, and inter-server communications can be seen as the sum of several parallel rings.

In such a case, we prove in the Appendix that 
for a large enough $n$, \ie for a large $\D$, 
it is highly likely that \textit{$D$'s degree is $k$}. As a result, there is at least one row or column with exactly $k$ nonzero elements, and therefore we can apply the lower bounds of \cref{sec:LBs} to some row/column $i$ using the parameter $k$. This applies to the benchmark workload in the evaluations.

\section{Evaluations}

\subsection{Settings}

We now want to evaluate our \name algorithm. These evaluations have two goals. First,
we evaluate \name on traffic matrices from \textit{AI training workloads}. To do so, we use both a relatively sparse and strongly skewed matrix from a \textit{GPT} model, as well as introduce a relatively dense and uniform matrix from a model with \textit{Mixture of Experts (MoE)}. A second goal is to compare \name with previous state-of-the-art algorithms that were designed for similar objectives, and understand the impact of different characteristics of $D$. To do so, we use a standard \textit{benchmark workload} that has already been used in the evaluations of several papers.

\T{AI training workload 1: GPT-3B.} Our first AI training workload results from a GPT model~\cite{GPT} with 3B parameters. Li et al.~\cite{AI-workloads} use Microsoft’s
DeepSpeed framework~\cite{deepspeed} to apply a hybrid parallelism that combines pipeline parallelism (PP), tensor parallelism (TP) and data parallelism (DP). They also adopt the default mapping approach of DeepSpeed in a platform of 32 RTX3090 GPUs. They obtain a $32 \times 32$ total traffic matrix $D$ that is 
strongly skewed and quite sparse. We then normalize $D$ to make it doubly substochastic, and add a small Gaussian noise to nonzero elements of standard deviation $\sigma=0.3\%$ as in the literature benchmark. Using random noise, we generate 50 random instances of $D$ then average our results.

\T{AI training workload 2: MoE.} We generate MoE routing traces by running three training iterations of the open-source Qwen2 MoE model (57B) on Megatron-LM across 64 NVIDIA H100 GPUs, organized as 8 HGX nodes (with $80\unit{GB}$ per GPU). We start from its released weights, use synthetic input tokens, and record the token-to-expert routing decisions. We use 64 experts, with each expert being placed on one of the 64 GPUs.
During each training iteration and each expert layer, we track communication traces between experts and accumulate each layer's traffic into a $64 \times 64$ demand matrix $D$, where each entry is the token count from source to destination. With 28 expert layers per iteration, we get $3\cdot 28 = 84$ distinct traffic demand matrices $D$, over which we average our results. We also later approximate this MoE workload using a Zipf distribution to study its sensitivity.

\T{Workload 3: Benchmark.} 
We use the standard benchmark in the field~\cite{solstice,eclipse,LESS}. It builds a $100 \times 100$ demand $D$ that consists of $m=16$ random flows per source port, including 4 large flows, which evenly split 70\% of the bandwidth, and 12 small flows, which evenly split 30\%, to reflect the sparse and skewed datacenter traffic. %
Each flow corresponds to a permutation matrix, and the sum of flows yields $D$. 
Finally, the nonzero demands are perturbed with small Gaussian noise of standard deviation $0.3\%$ of the link bandwidth. Using random  noise, we generate 50 random instances of $D$ then average our results.

\T{Algorithms.} We want to evaluate our \name algorithm against the best state-of-the-art optical switch scheduling algorithms. Unfortunately, most algorithms in the literature consider a single optical switch and cannot be applied to scheduling across $s$ parallel optical switches. We compare \name against several algorithms:

\Ts{(1) \base.} The closest to our work is
\less~\cite{LESS}, which considers scheduling a demand matrix $D$ across $s$ parallel optical switches by splitting $D$ into $s$ sub-workload matrices $D_1,  \ldots, D_s$, each $D_i$ to be independently scheduled on switch $i$. The splitting objective is to maximize the \textit{sparsity} in these sub-workload matrices by minimizing the sum of nonzero elements across $D_1, \ldots, D_s$. While \less uses a partial-reconfiguration scheduling algorithm intended for partially-reconfigurable switches,
we adopt its sparsity-based approach as our \base for comparisons.
For an apples-to-apples comparison, we also use our \decomp algorithm to schedule the sub-workload matrices in \base, since \less does not provide a similar decomposition algorithm.

We further introduce three algorithms that only replace the \decomp step in \name while keeping the other steps, thus obtaining three \name variants:

\Ts{(2) \name (\eclipse).} This variant relies on  \eclipse~\cite{eclipse}, a widely used algorithm 
in decompositions with reconfiguration delays, to replace the \decomp step in \name.

\Ts{(3) \name (\birkhoff).}  This variant relies on \birkhoff~\cite{valls2021birkhoff},  a novel algorithm in contexts where reconfiguration delay is not considered. The algorithm makes an approximate decomposition that is at most $\epsilon$ off from the true demand $D$'s Frobenius norm. 
Since we want to schedule all traffic, $\epsilon$ is set to be near $0$ (about $10^{-12}$). Additionally, \birkhoff is designed for doubly stochastic matrices; however, our workload mostly consists of doubly substochastic matrices. In such cases, we also use von Neumann's algorithm~\cite{vonneumann1953,chang2000birkhoff} that adds demand to $D$ to make it doubly stochastic, before applying \birkhoff.

\Ts{(4) \name (\wwfa).} Chronos~\cite{renganathan2025chronos} has recently suggested using \wwfa (Wrapped Wave Front Arbiter) as a greedy MWM approximation. We also examine a \name version that uses \wwfa for the \decomp step.

\T{Implementation and Runs.} 
For the Hungarian algorithm, we implemented an efficient variant called the Jonker and Volgenant algorithm that runs faster in practice~\cite{jonker1999linear, jonker,crouse}.
We conduct our experiments on a computer with a 3.7 GHz AMD Ryzen Threadripper 3970X Processor and 128\,GB of memory, running on Ubuntu 20.04.6 LTS. 
The runtimes for \name 
are from under 1\,ms to tens of ms.%
\footnote{Datacenter switches typically use specialized network processors or ASICs, so \name runtimes can be much faster in practice.}
Each datapoint is the average of 50 runs with random matrices, except for MoE, where it is the average over the 84 measured matrices. %

\subsection{AI Workloads}

\begin{figure}
    \centering
    \includegraphics[width=0.8\linewidth]{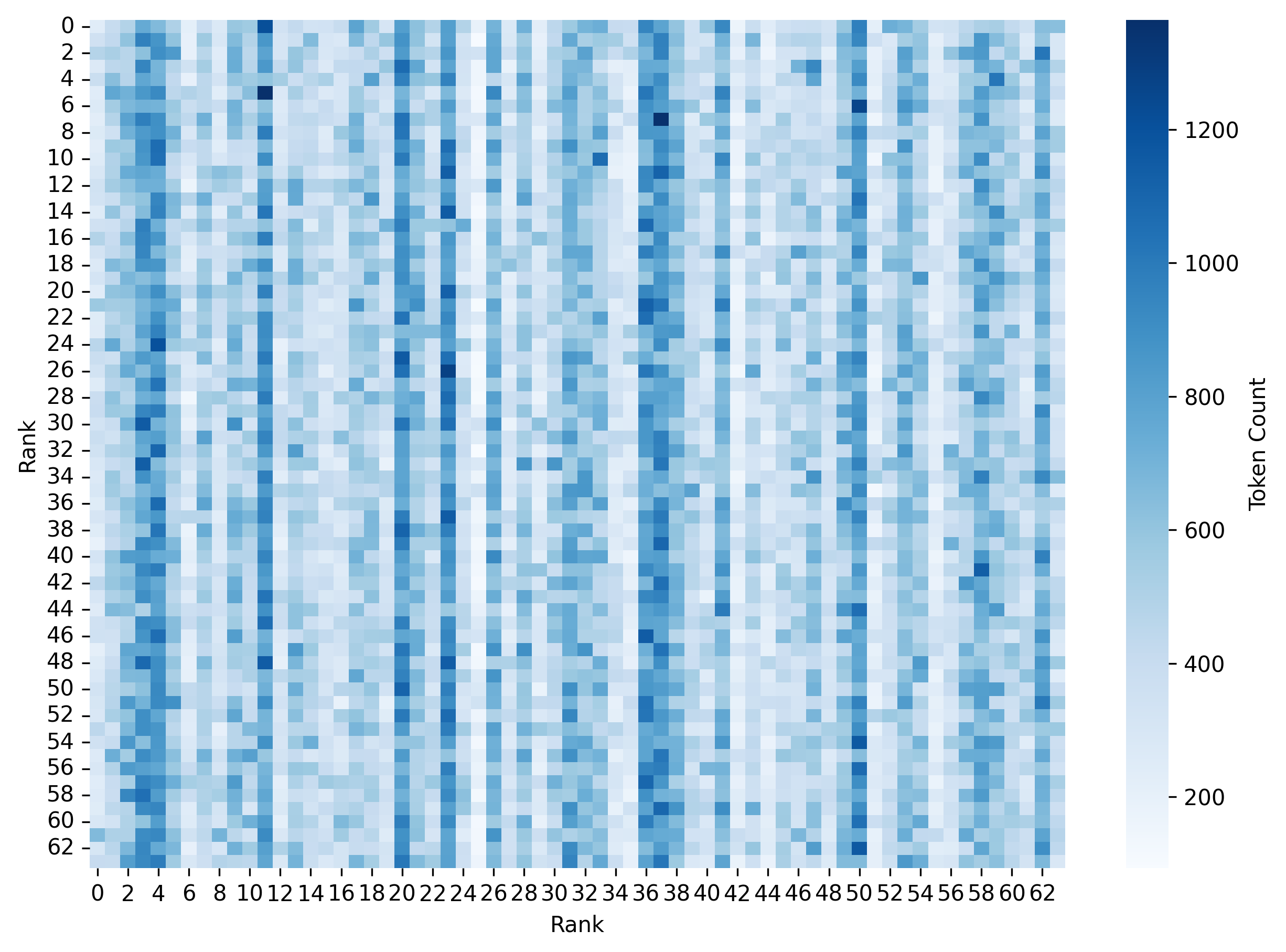}
    \caption{MoE traffic matrix heatmap from 64 sender GPUs (rows) to 64 receiver GPUs (columns) for the first layer in the first iteration of training.
    }
    \label{fig:moematrix}
\end{figure}

\Cref{fig:moematrix} shows the distribution of traffic during the first iteration and first expert layer
of the Qwen model. Each GPU expert sends out $32768$ tokens (row sums of the demand matrix). Each token consists of $2$ bytes of data (BF16 encoding). Its traffic is relatively dense. It is slightly non-uniform between different destination experts, as we can see some heavier columns, 
but more dense and uniform than the GPT pattern, and slightly more uniform than some other smaller examples in the literature~\cite{liao2025mixnet,renganathan2025chronos,gandhi2024moetion},~\cite{cohen2026incast}.

\begin{figure*}[!t] %
\centering
\begin{subfigure}{0.48\linewidth}
  \includegraphics[width=\linewidth]{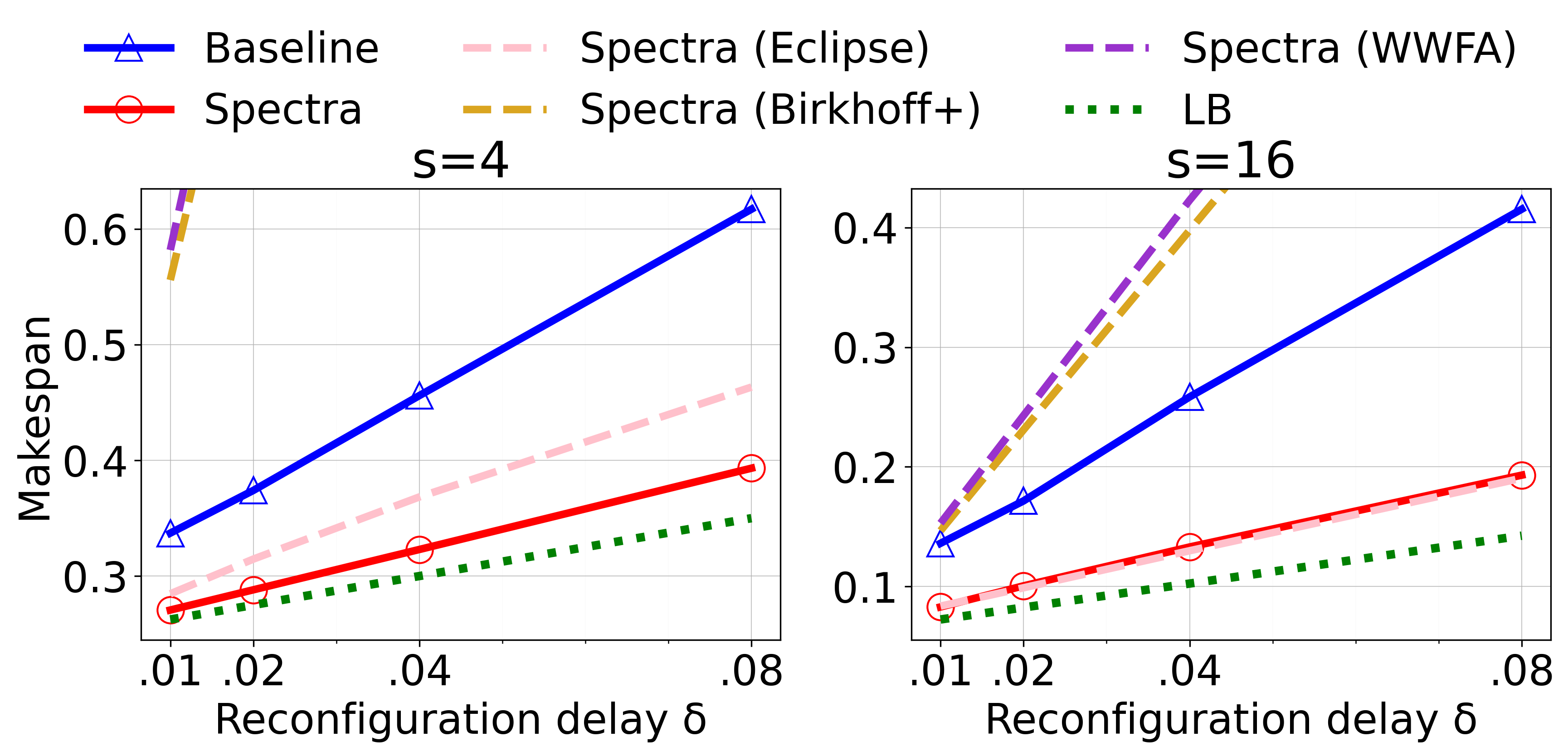} 
    \caption{GPT with $n=32$.}
    \label{fig:wwfa_gpt}
\end{subfigure}
\hfill
\begin{subfigure}{0.48\linewidth}
\includegraphics[width=\linewidth]{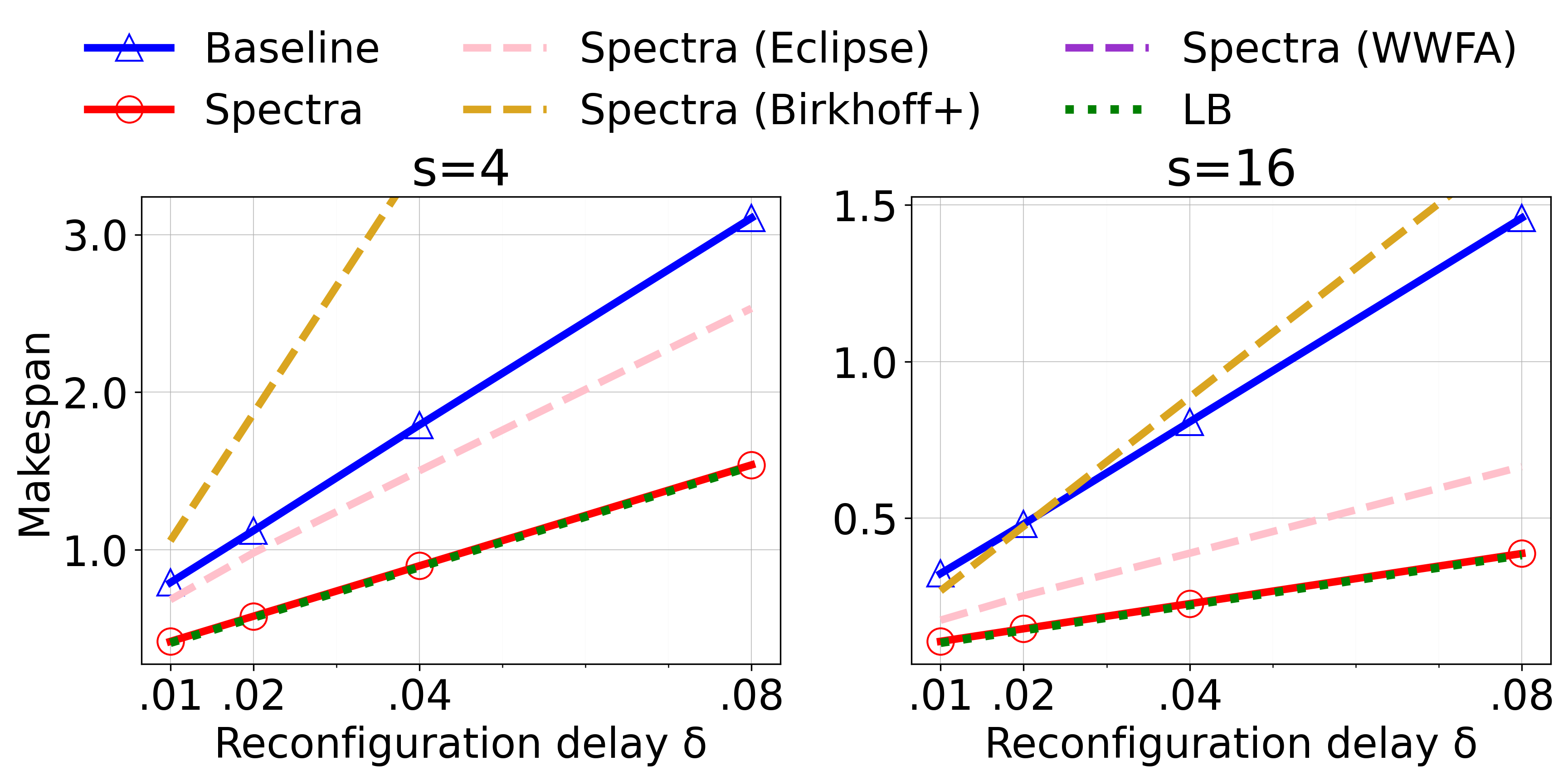}
    \caption{MoE with $n=64$.
    }
    \label{fig:wwfa_qwen}
\end{subfigure}
\caption{Performance with \textit{AI training workloads} 
given varying reconfiguration delay $\delta$, with (a)~GPT and (b)~MoE workloads, using two different numbers $s$ of OCSes. 
} 
\label{fig:AI-WWFA}
\SmallReduceVSpace %
\end{figure*}

\begin{figure*}[!t] %
\centering
\begin{subfigure}{0.48\linewidth}
  \includegraphics[width=\linewidth]{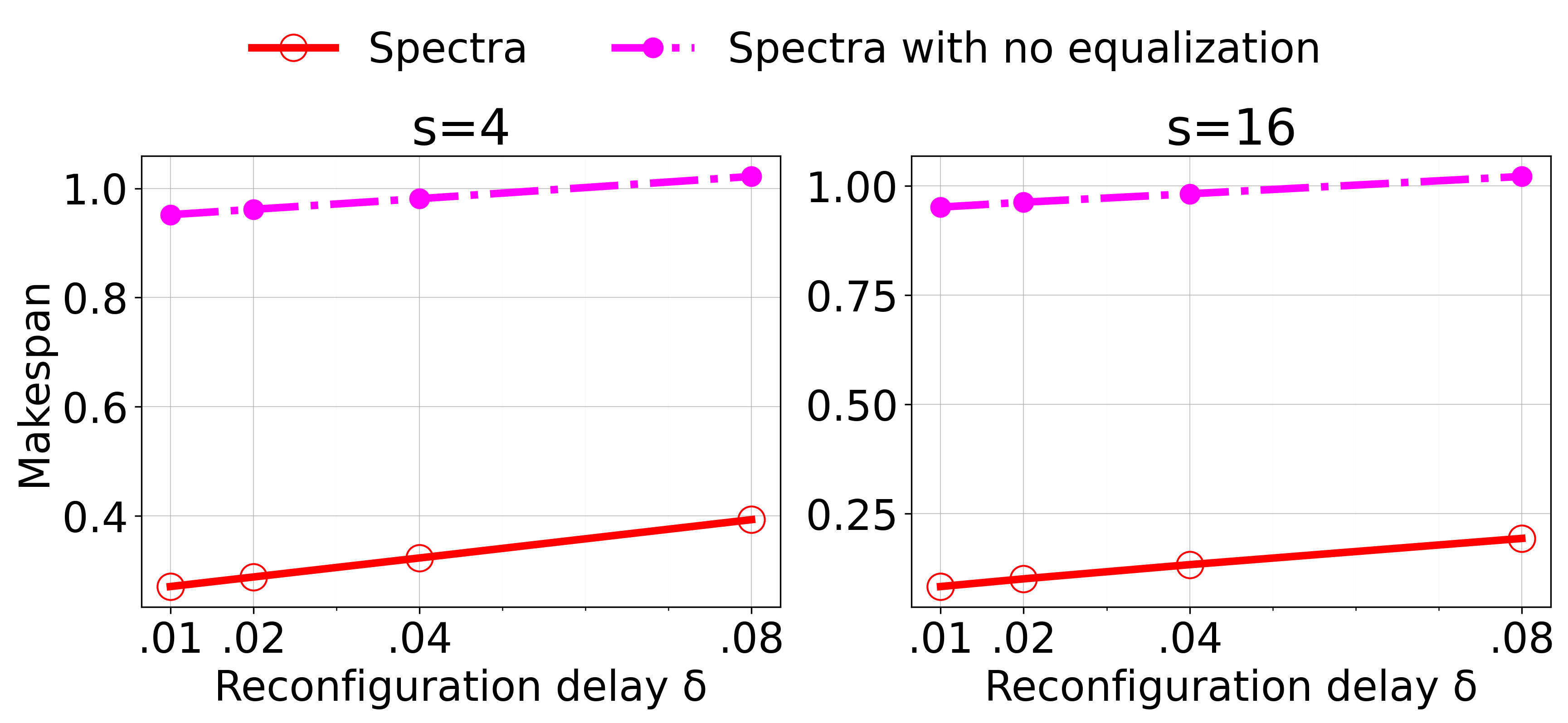}
    \caption{GPT with $n=32$.}
    \label{fig:ai3.4}
\end{subfigure}
\hfill
\begin{subfigure}{0.48\linewidth}
    \includegraphics[width=\linewidth]{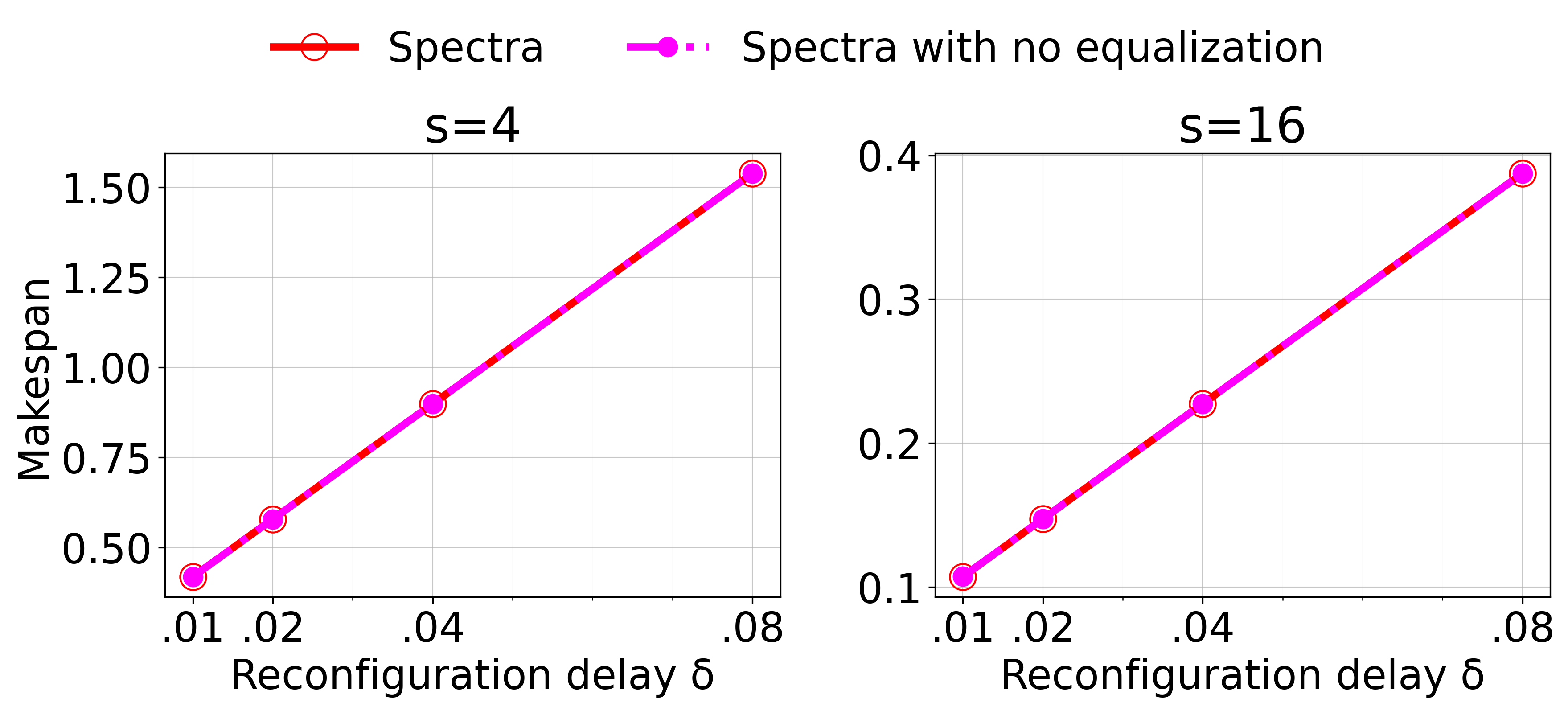}
    \caption{MoE with $n=64$.}
    \label{fig:ai3.1}
\end{subfigure}
\caption{\textit{Sensitivity to equalization} with \textit{AI training workloads}, comparing \name with and without equalization. %
}
\label{fig:AI-eq}
\SmallReduceVSpace %
\end{figure*}

\Cref{fig:AI-WWFA} presents the results for the sparse skewed GPT workload (\cref{fig:wwfa_gpt}) and for the dense near-uniform MoE workload (\cref{fig:wwfa_qwen}). 
It also shows our derived lower bound (\cref{sec:LBs}). First,\textit{ \name appears close to optimal}, since it is close to the lower bound (LB), especially with the MoE workload where it is indistinguishable, maybe because it can focus on the few columns of highest-weight destination experts (\cref{fig:moematrix}). Moreover, in all cases,  \name far outperforms  \base. In fact, \name's schedule is $1.4\times$
shorter than \base on average on GPT traffic, and $2.7\times$ 
shorter on average on MoE. Specifically, not only does \name perform much better for low reconfiguration cost $\delta$, but also its makespan grows slower as $\delta$ grows. This is because \base tries to optimize for a minimal number of nonzero elements to each switch's sub-workload demand matrix, but this optimization does not directly minimize the number of reconfigurations and does not necessarily translate into a lower makespan. Thus, this sparsity objective does not appear to fit our settings.

Using \eclipse for the \decomp step achieves close results with the GPT workload due to its well-structured and nearly doubly stochastic matrix, but performs much worse with the denser and substochastic MoE matrix. That is, the \eclipse-based variant of \name performs $1.1\times$ worse on average for GPT, and $1.7\times$ worse on average for MoE. %
Using \birkhoff for the \decomp step is actually worse than \base for the GPT and MoE workloads, performing worse than \base by $2.1\times$ for the GPT workload, and worse than \base by $1.4\times$ on average for the MoE workload. It likely does not perform well because it does not take into account reconfiguration delay, and given a low $\epsilon$, \birkhoff will decompose $D$ into many permutations with very low weight compared to the reconfiguration delay. Using \wwfa for the \decomp step attains the worst performance. In \Cref{fig:wwfa_qwen}, the makespan for \wwfa is so large that it had to be cropped out of the graph. \wwfa is worse than \base by $2.3\times$ on average for the GPT workload, and by $14.5\times$ on average for the MoE workload. This is likely because, in addition to not considering reconfiguration delays, \wwfa is also greedy and does not efficiently decompose demand $D$ into a small number of permutations.
The rest of this section will thus exclude \wwfa and \birkhoff from the graphs for clarity of the rest of the data. 
In summary, the results show that \name performs well on both sparse and dense matrices, which highlights its flexibility.

\Cref{fig:AI-eq} analyzes the sensitivity of \name to its equalization component. We can see that when we need to schedule the dense and near-uniform MoE traffic with many small elements (\cref{fig:ai3.1}), it is relatively easy to spread the traffic, and therefore equalization does not help. However, as we 
use the GPT traffic with large elements (\cref{fig:ai3.4}), the elements need to be split, and \textit{equalization makes a significant difference}.

\begin{figure*}[!t] %
\centering
\begin{subfigure}{0.48\linewidth}
  \includegraphics[width=\linewidth]{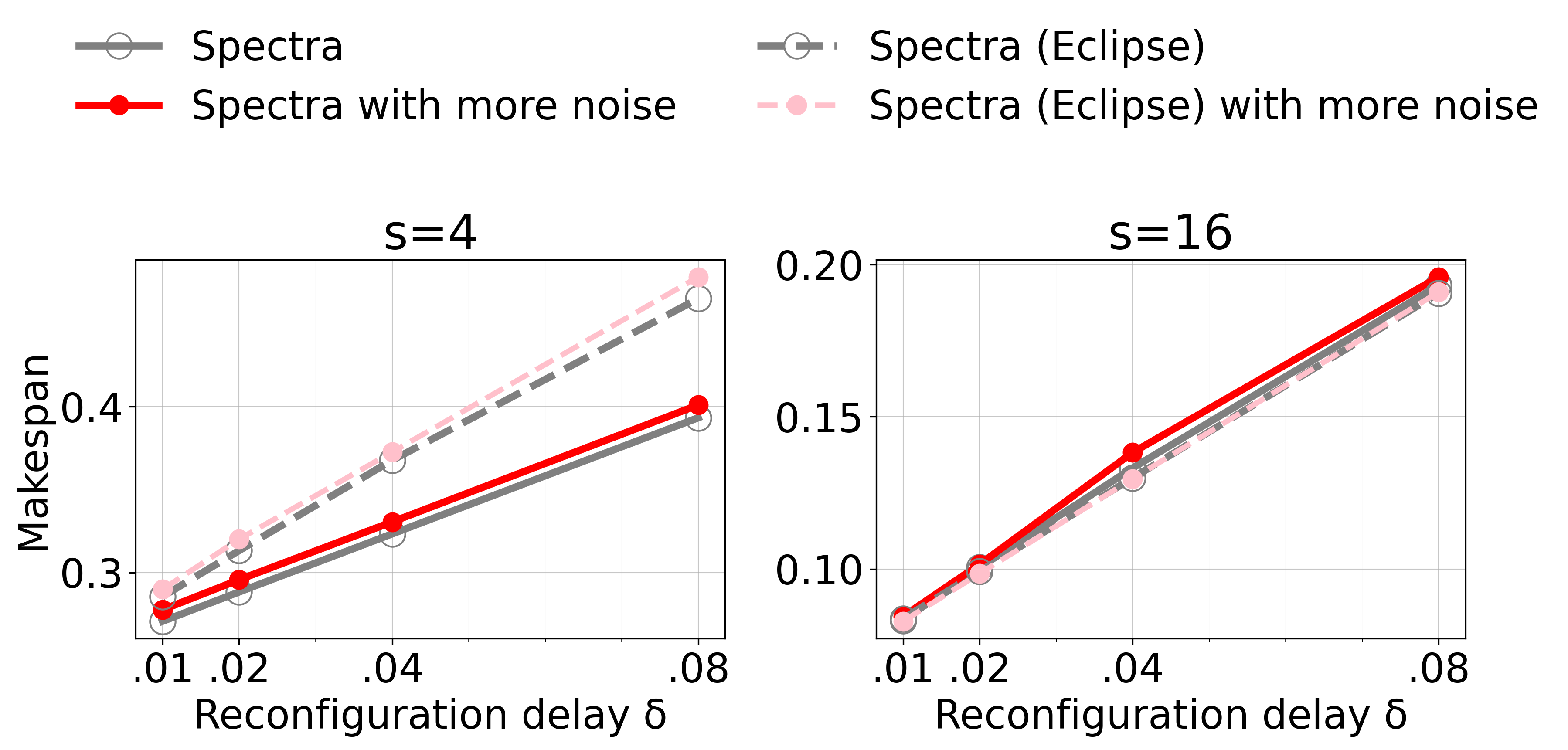}
    \caption{GPT with $n=32$.}
    \label{fig:ai9.4}
\end{subfigure}
\hfill
\begin{subfigure}{0.48\linewidth}
    \includegraphics[width=\linewidth]{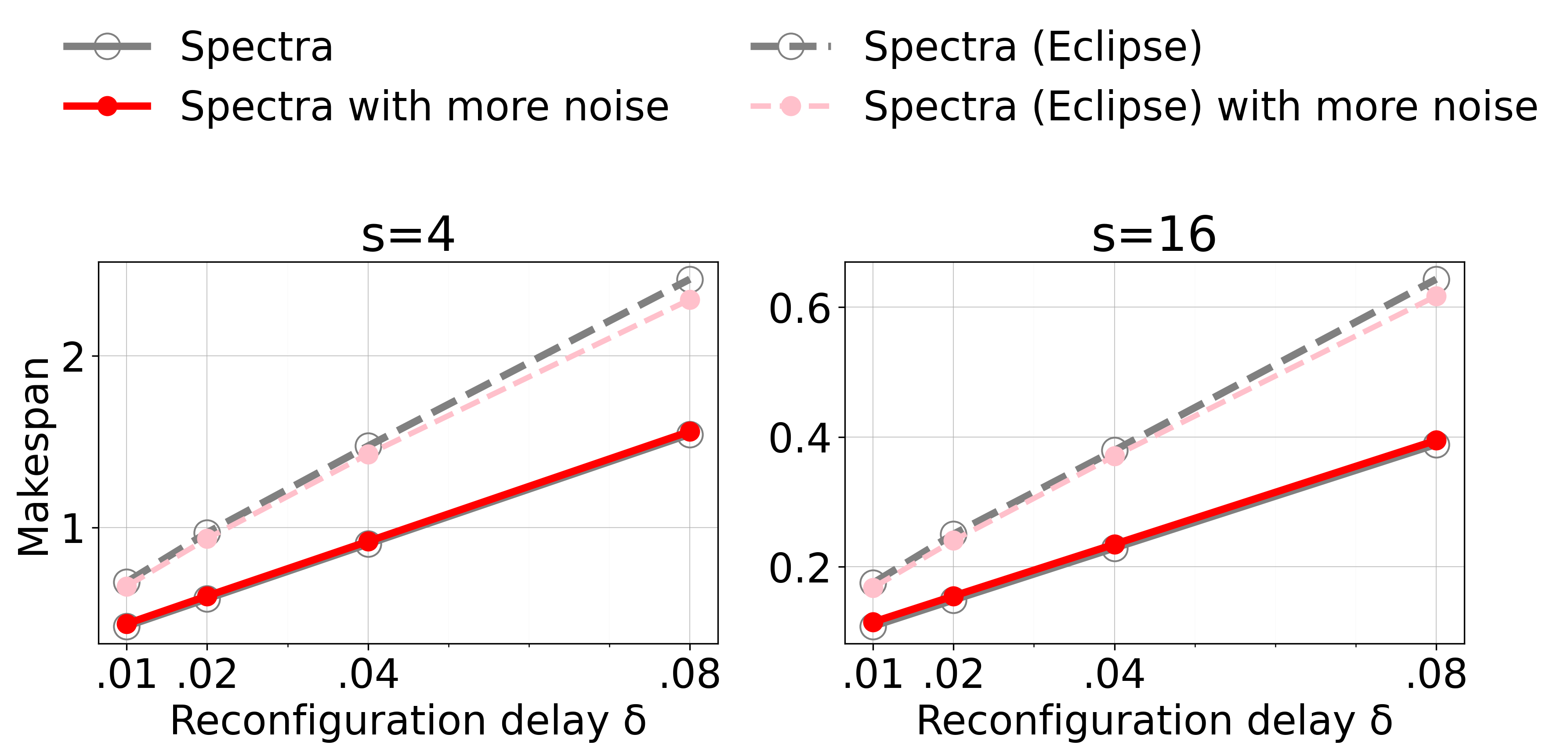}
    \caption{MoE with $n=64$.}
    \label{fig:ai9.1}
\end{subfigure}
\caption{\textit{Sensitivity to noise} with \textit{AI training workload}: Comparison of \name variants with noise standard deviations $0.3\%$ and $1\%$.
}
\label{fig:AI-noise}
\SmallReduceVSpace %
\end{figure*}

\Cref{fig:AI-noise} presents the impact of added noise on the performance of \name and \name with \eclipse in the AI training workloads. It plots the result of having a small noise added to all non-zero elements (of standard deviation $\sigma=0.3\%$, plotted in gray) \vs having a larger noise ($\sigma=1\%$, plotted with usual colors). The 0.3\% noise is already added in the initial GPT workload, while we add it specifically in the MoE workload for the comparison.
The GPT initial traffic matrix is more regular and close to being doubly stochastic, 
and therefore we can see that noise has a slightly negative impact, especially for $s=4$. On the contrary, MoE traffic is strongly substochastic, so the noise can in fact make it less skewed and improve the performance.

\begin{figure}
    \centering
    \includegraphics[width=\linewidth]{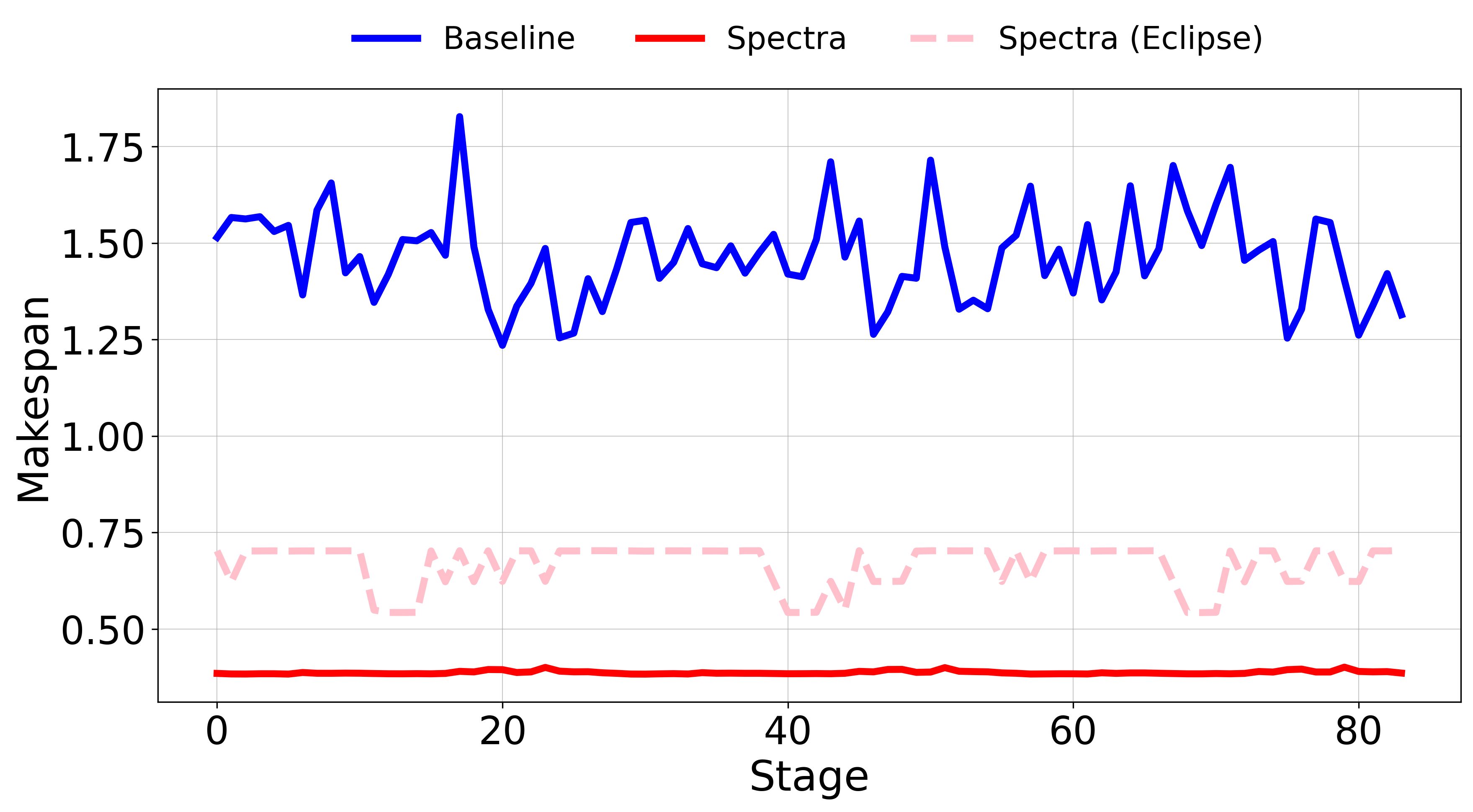}
    \caption{\textit{Sensitivity to traffic change} with AI training workload (MoE) across 84 stages of training.
    }
    \label{fig:stages_merged}
\end{figure}

\Cref{fig:stages_merged} analyzes the sensitivity of the algorithms to the evolution of traffic during the 84 stages of training the MoE model. In addition to having a consistently lower makespan than the other approaches, \name also has the lowest variance and the most predictability versus the other approaches.

\begin{figure*}
    \centering
    \includegraphics[width= \linewidth]{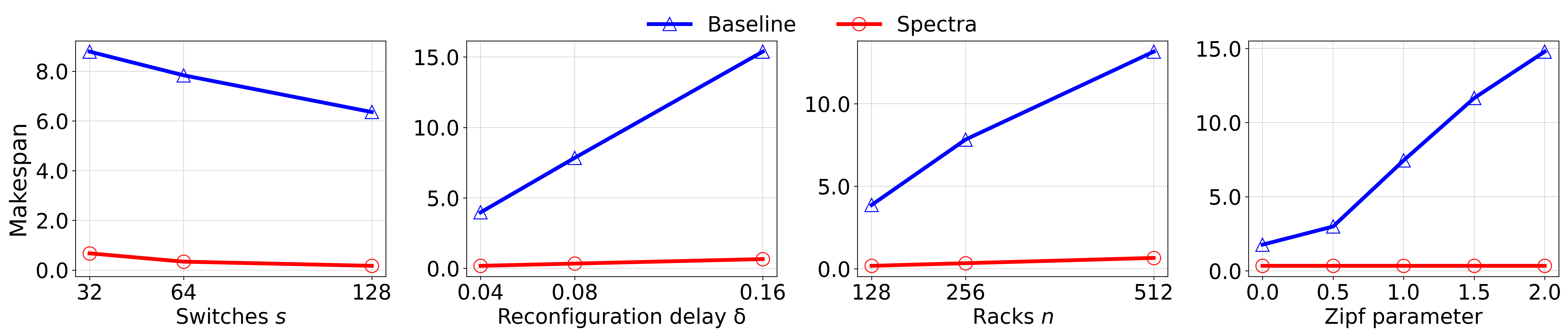}
    \caption{\textit{Sensitivity to scaling} with Zipf workload.}
    \label{fig:zipf_horizontal}
\end{figure*}

\Cref{fig:zipf_horizontal} analyzes the sensitivity of the algorithms to scaling the number $n$ of racks, the reconfiguration delay $\delta$, and the number $s$ of switches. Since we want to scale to larger $n$, we approximate the MoE traffic matrices (which use $n=64$) by using a synthetic traffic with a Zipf distribution of parameter 1. We also vary this parameter.
We can see that as we increase the number of racks or the reconfiguration delay, \name and \base have a proportional near-linear increase in makespan, with \name achieving a much better performance. Likewise, \name's makespan clearly outperforms that of \base as the number $s$ of switches or the Zipf parameter vary. 

\subsection{Benchmark Workload}

\begin{figure}
    \centering
    \includegraphics[width=\linewidth]{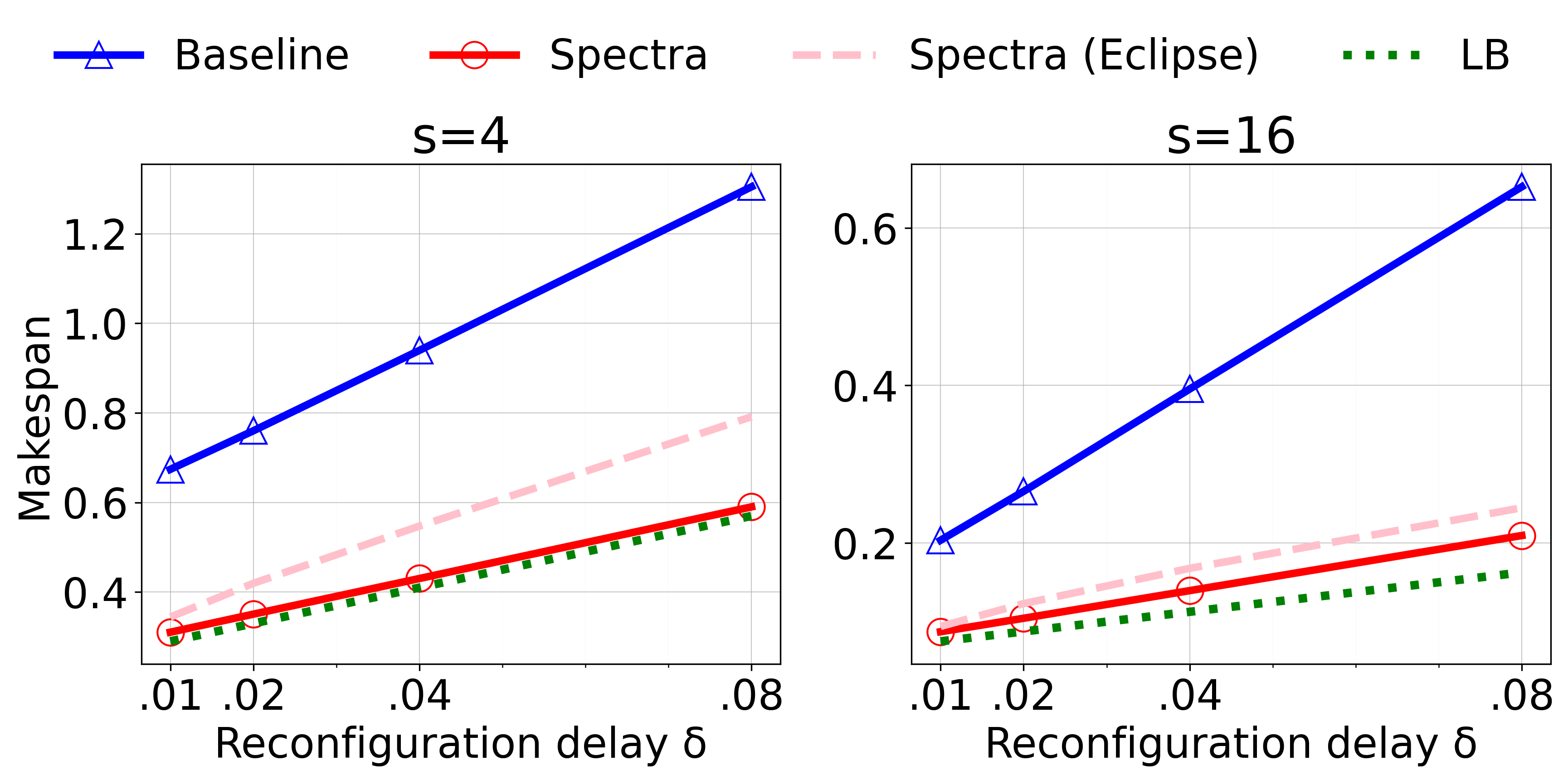}
    \caption{\textit{Benchmark workload}. 
    }
    \label{fig:syn1}
\SmallReduceVSpace \end{figure}

\Cref{fig:syn1} compares \name against alternatives on the benchmark workload. 
\name vastly outperforms \base, with a $2.4\times$ 
shorter schedule on average. Using \eclipse for the  \decomp step always yields worse results, and \name performs better than \name (\eclipse) with a $1.2\times$
shorter schedule on average. 
\name also appears quite close to the lower bound, and therefore, to optimality.

\begin{figure}
    \centering
    \includegraphics[width=\linewidth]{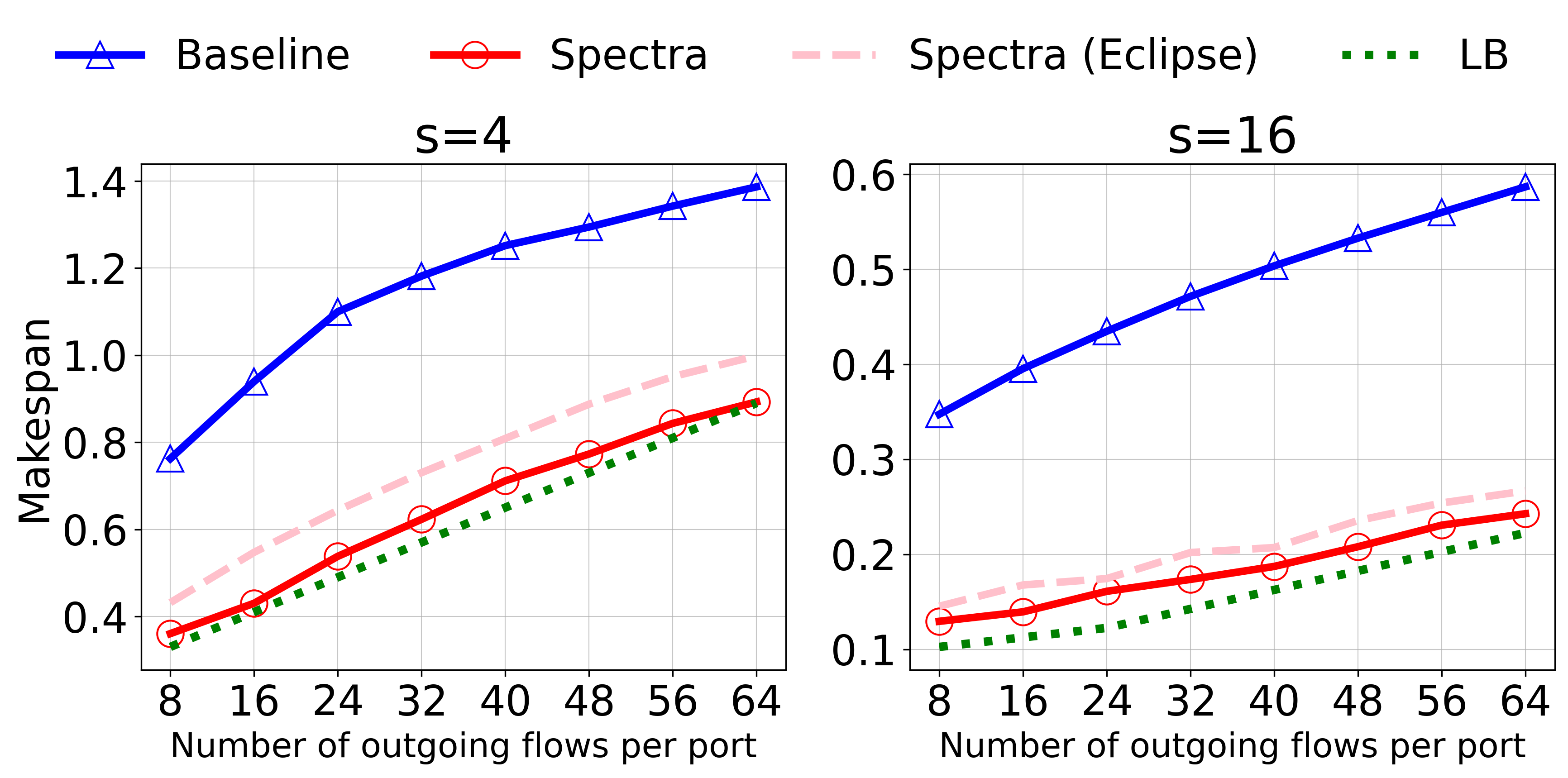}
    \caption{\textit{Sensitivity to sparsity} with \textit{benchmark workload}, with varying number of flows. %
    }
    \label{fig:syn2}
\end{figure}

\Cref{fig:syn2} analyzes the sensitivity of the algorithms to a varying number $m$ of outgoing flows per port, 
and therefore to a varying sparsity of the matrix $D$, given reconfiguration delay $\delta = 0.04$.  \name is flexible enough to adapt to these differing sparsity patterns, again vastly outperforming \base. Using an \eclipse decomposition step never helps \name.

\section{Conclusion}

In this paper, we showed how the increasing network demands in AI datacenters
could be serviced by a set of parallel OCSes, in which we want to minimize the makespan. We then explained how \name can schedule a demand matrix $D$ over $s$ parallel switches using three steps: \decomp, \sched and \equal. 
Our evaluation of \name shows that it vastly outperforms \base by 
an average factor of $1.4\times$ on GPT AI workloads, $2.7\times$ on MoE AI workloads, and $2.4\times$ on standard benchmarks.

\section*{Acknowledgment}
The authors would like to thank Ori Cohen for providing the MoE measurements, and Mark Silberstein and Jose Yallouz for their assistance in coordinating the necessary resources.

\appendix

\T{Sum of $k$ random weighted permutations.} We focus on the case where the $n \times n$ demand matrix $\D$ is the sum of $k$ random weighted permutations, as in the benchmark workload in the evaluations (before noise is added). We want to show that 
for a large enough $n$, it is highly likely that \textit{$D$'s degree is $k$}, meaning there is at least one row or column with exactly $k$ nonzero elements, and therefore we can apply the lower bounds of \cref{sec:LBs} to some row/column $i$ using the parameter $k$. We start by proving a property of a single row or column of $\D$.

\begin{proposition}
If $\D$ is defined as the sum of $k$ random weighted permutations, then the probability $p$ that there are exactly $k$ nonzero elements in a given row or column of $\D$ is 
\be
p=\frac{n!}{(n-k)! \cdot n^k}
\ee
\end{proposition}
\bp
Consider $k$ colored balls that each randomly choose one of $n$ bins. $p$ is the probability that they end up in distinct bins. It is the division of the number $n \cdot (n-1) \cdot \dots \cdot (n-k+1) = \frac{n!}{(n-k)!}$ of possible choices for the placement of the balls such that they get in distinct bins, by the total number $n^k$ of possible placements.
\ep

We now build a model that approximates the probability that $\D$ has degree $k$.
This model considers all rows and columns of $\D$ as i.i.d. for this approximation.

\begin{figure}[!t] %
\centering
\begin{subfigure}{0.48\linewidth}
\includegraphics[width=\columnwidth]{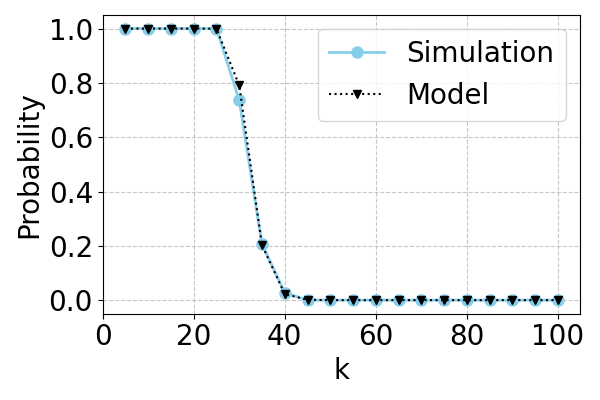}
  \caption{$k$ varies, given $n=100$}
  \label{fig:model1}
\end{subfigure}
\hfill
\begin{subfigure}{0.48\linewidth}
  \includegraphics[width=\columnwidth]{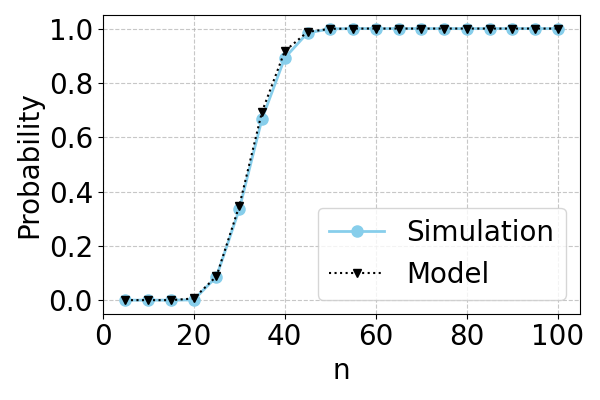}
  \caption{$n$ varies, given $k=16$}
  \label{fig:model2}
\end{subfigure}
\caption{Probability that 
an $n\times n$  matrix $\D$ that is the sum of $k$ random weighted permutations also has degree $k$.}
\label{fig:cbiq}
\ReduceVSpace %
\end{figure}

\begin{proposition}
If $\D$ is defined as the sum of $k$ random weighted permutations, then the probability that  its degree is $k$ 
(and therefore the lower bounds for $i$ 
hold with $k$ nonzero elements) 
can be approximated as:
\be
1- \para{1-\frac{n!}{(n-k)! \cdot n^k}}^{2n}.
\ee
\end{proposition}
\bp
We approximate the $2n$ rows and  columns as i.i.d., with a probability $p$ that any given row or column has $k$ nonzero elements. Then the  
probability that at least one row or column has $k$ nonzero elements is $
1-(1-p)^{2n}$. %
\ep

\Cref{fig:model1} illustrates the simulated probability that the degree of $D$ is $k$ 
in a $100\times 100$  matrix $\D$ that is the sum of $k$ random weighted permutations. It
shows that the approximation is close. %
\cref{fig:model2} shows the same probability for a sum of 16 permutations, as $D$'s size $n$ varies. Again, the approximation holds well. Moreover, for $n$ beyond 50, it is highly likely that its degree will be 16.

 \bibliographystyle{IEEEtran}
\bibliography{mybib}

\begin{thebibliography}{10}
\providecommand{\url}[1]{#1}
\csname url@samestyle\endcsname
\providecommand{\newblock}{\relax}
\providecommand{\bibinfo}[2]{#2}
\providecommand{\BIBentrySTDinterwordspacing}{\spaceskip=0pt\relax}
\providecommand{\BIBentryALTinterwordstretchfactor}{4}
\providecommand{\BIBentryALTinterwordspacing}{\spaceskip=\fontdimen2\font plus
\BIBentryALTinterwordstretchfactor\fontdimen3\font minus
  \fontdimen4\font\relax}
\providecommand{\BIBforeignlanguage}[2]{{%
\expandafter\ifx\csname l@#1\endcsname\relax
\typeout{** WARNING: IEEEtran.bst: No hyphenation pattern has been}%
\typeout{** loaded for the language `#1'. Using the pattern for}%
\typeout{** the default language instead.}%
\else
\language=\csname l@#1\endcsname
\fi
#2}}
\providecommand{\BIBdecl}{\relax}
\BIBdecl

\bibitem{liang-net}
K.~Liang, L.~Qiao, I.~Keslassy, and B.~Lin, ``Scheduling parallel optical
  switches in datacenters,'' \emph{IEEE Networking Letters}, vol.~8, pp.
  149--153, 2026.

\bibitem{liao2025mixnet}
X.~Liao, Y.~Sun, H.~Tian, X.~Wan, Y.~Jin, Z.~Wang, Z.~Ren, X.~Huang, W.~Li,
  K.~F. Tse \emph{et~al.}, ``{MixNet}: A runtime reconfigurable
  optical-electrical fabric for distributed mixture-of-experts training,'' in
  \emph{{ACM SIGCOMM}}, 2025, pp. 554--574.

\bibitem{meta}
A.~Gangidi, R.~Miao, S.~Zheng, S.~J. Bondu, G.~Goes, H.~Morsy, R.~Puri,
  M.~Riftadi, A.~J. Shetty, J.~Yang \emph{et~al.}, ``{RDMA} over {Ethernet} for
  distributed training at {Meta} scale,'' in \emph{{ACM SIGCOMM}}, 2024, pp.
  57--70.

\bibitem{Stellar}
J.~Lu, J.~Gao, F.~Feng, Z.~He, M.~Zheng, K.~Liu, J.~He, B.~Liao, S.~Xu, K.~Sun
  \emph{et~al.}, ``{Alibaba Stellar}: A new generation {RDMA} network for cloud
  {AI},'' in \emph{{ACM SIGCOMM}}, 2025, pp. 453--466.

\bibitem{alibaba}
K.~Qian, Y.~Xi, J.~Cao, J.~Gao, Y.~Xu, Y.~Guan, B.~Fu, X.~Shi, F.~Zhu, R.~Miao
  \emph{et~al.}, ``Alibaba {HPN}: A data center network for large language
  model training,'' in \emph{Proceedings of the ACM SIGCOMM 2024 Conference},
  2024, pp. 691--706.

\bibitem{feng2025railx}
Y.~Feng, T.~Chen, Y.~Wei, S.~Shen, S.~Wang, W.~Li, K.~Ma, and T.~Hoefler,
  ``{RailX}: a flexible, scalable, and low-cost network architecture for
  hyper-scale {LLM} training systems,'' \emph{arXiv preprint arXiv:2507.18889},
  2025.

\bibitem{eclipse}
S.~Bojja~Venkatakrishnan, M.~Alizadeh, and P.~Viswanath, ``Costly circuits,
  submodular schedules and approximate carath{\'e}odory theorems,''
  \emph{Queueing Systems}, vol.~88, no.~3, pp. 311--347, 2018.

\bibitem{solstice}
H.~Liu \emph{et~al.}, ``Scheduling techniques for hybrid circuit/packet
  networks,'' in \emph{{ACM CoNEXT}}, 2015, pp. 1--13.

\bibitem{farrington2010helios}
N.~Farrington \emph{et~al.}, ``Helios: a hybrid electrical/optical switch
  architecture for modular data centers,'' in \emph{{ACM SIGCOMM}}, vol.~40,
  no.~4, 2010, pp. 339--350.

\bibitem{valls2021birkhoff}
V.~Valls, G.~Iosifidis, and L.~Tassiulas, ``Birkhoff’s decomposition
  revisited: Sparse scheduling for high-speed circuit switches,''
  \emph{IEEE/ACM Transactions on Networking}, vol.~29, no.~6, pp. 2399--2412,
  2021.

\bibitem{LESS}
L.~Liu, J.~J. Xu, and M.~Singh, ``{LESS}: A matrix split and balance algorithm
  for parallel circuit (optical) or hybrid data center switching and more,'' in
  \emph{IEEE/ACM Utility and Cloud Computing (UCC)}, 2019, pp. 187--197.

\bibitem{negotiator}
C.~Liang \emph{et~al.}, ``{NegotiaToR}: towards a simple yet effective
  on-demand reconfigurable datacenter network,'' in \emph{{ACM SIGCOMM}}, 2024.

\bibitem{zu2024resiliency}
Y.~Zu, A.~Ghaffarkhah, H.-V. Dang, B.~Towles, S.~Hand, S.~Huda, A.~Bello,
  A.~Kolbasov, A.~Rezaei, D.~Du \emph{et~al.}, ``Resiliency at scale: Managing
  {Google's} {TPUv4} machine learning supercomputer,'' in \emph{Usenix NSDI},
  2024, pp. 761--774.

\bibitem{unlocking}
F.~De~Marchi, J.~Li, Y.~Zhang, W.~Bai, and Y.~Xia, ``Unlocking superior
  performance in reconfigurable data center networks with credit-based
  transport,'' in \emph{{ACM SIGCOMM}}, 2025, pp. 842--860.

\bibitem{vermilion}
V.~Addanki \emph{et~al.}, ``Vermilion: A traffic-aware reconfigurable optical
  interconnect with formal throughput guarantees,'' in \emph{Usenix NSDI},
  2027.

\bibitem{sirius}
H.~Ballani, P.~Costa, R.~Behrendt, D.~Cletheroe, I.~Haller, K.~Jozwik,
  F.~Karinou, S.~Lange, K.~Shi, B.~Thomsen \emph{et~al.}, ``Sirius: A flat
  datacenter network with nanosecond optical switching,'' in \emph{{ACM
  SIGCOMM}}, 2020, pp. 782--797.

\bibitem{topoopt}
W.~Wang, M.~Khazraee, Z.~Zhong, M.~Ghobadi, Z.~Jia, D.~Mudigere, Y.~Zhang, and
  A.~Kewitsch, ``{TopoOpt}: Co-optimizing network topology and parallelization
  strategy for distributed training jobs,'' in \emph{Usenix NSDI}, 2023, pp.
  739--767.

\bibitem{hamdi}
X.~Li and M.~Hamdi, ``On scheduling optical packet switches with
  reconfiguration delay,'' \emph{{IEEE JSAC}}, vol.~21, no.~7, pp. 1156--1164,
  2003.

\bibitem{schrijver2003combinatorial}
A.~Schrijver, \emph{Combinatorial Optimization: Polyhedra and Efficiency}, ser.
  Algorithms and Combinatorics.\hskip 1em plus 0.5em minus 0.4em\relax Berlin:
  Springer, 2003, vol.~24, no.~2.

\bibitem{ahuja1993network}
R.~K. Ahuja, T.~L. Magnanti, and J.~B. Orlin, \emph{Network Flows: Theory,
  Algorithms, and Applications}.\hskip 1em plus 0.5em minus 0.4em\relax
  Prentice Hall, 1993.

\bibitem{berge2024approximation}
P.~Berg{\'e}, M.~Chaikovskaia, J.-P. Gayon, and A.~Quilliot, ``Approximation
  algorithm for job scheduling with reconfigurable resources,'' in
  \emph{International Symposium on Combinatorial Optimization}, 2024, pp.
  263--276.

\bibitem{gabow2018data}
H.~N. Gabow, ``Data structures for weighted matching and extensions to
  b-matching and f-factors,'' \emph{ACM Transactions on Algorithms (TALG)},
  vol.~14, no.~3, pp. 1--80, 2018.

\bibitem{jonker1999linear}
R.~Jonker and A.~Volgenant, ``Linear assignment procedures,'' \emph{European
  journal of operational research}, vol. 116, no.~1, pp. 233--234, 1999.

\bibitem{jonker}
------, ``A shortest augmenting path algorithm for dense and sparse linear
  assignment problems,'' \emph{Computing}, vol.~38, no.~4, pp. 325--340, 1987.

\bibitem{rajbhandari2020zero}
S.~Rajbhandari, J.~Rasley, O.~Ruwase, and Y.~He, ``{ZeRO}: Memory optimizations
  toward training trillion parameter models,'' in \emph{SC20: International
  Conference for High Performance Computing, Networking, Storage and
  Analysis}.\hskip 1em plus 0.5em minus 0.4em\relax IEEE, 2020, pp. 1--16.

\bibitem{zhao2023pytorch}
Y.~Zhao, A.~Gu, R.~Varma, L.~Luo, C.-C. Huang, M.~Xu, L.~Wright,
  H.~Shojanazeri, M.~Ott, S.~Shleifer \emph{et~al.}, ``{PyTorch FSDP}:
  Experiences on scaling fully sharded data parallel,'' \emph{Proceedings of
  the VLDB Endowment}, vol.~16, no.~12, pp. 3848--3860, 2023.

\bibitem{GPT}
A.~Radford, J.~Wu, R.~Child, D.~Luan, D.~Amodei, I.~Sutskever \emph{et~al.},
  ``Language models are unsupervised multitask learners,'' \emph{OpenAI blog},
  vol.~1, no.~8, p.~9, 2019.

\bibitem{AI-workloads}
W.~Li, X.~Liu, Y.~Li, Y.~Jin, H.~Tian, Z.~Zhong, G.~Liu, Y.~Zhang, and K.~Chen,
  ``Understanding communication characteristics of distributed training,'' in
  \emph{Proceedings of the 8th Asia-Pacific Workshop on Networking}, 2024, pp.
  1--8.

\bibitem{deepspeed}
``{Megatron-DeepSpeed},''
  \url{https://github.com/microsoft/Megatron-DeepSpeed}.

\bibitem{vonneumann1953}
J.~von Neumann, ``A certain zero-sum two-person game equivalent to the optimal
  assignment problem,'' in \emph{Contributions to the Theory of Games {II}},
  ser. Annals of Mathematics Studies.\hskip 1em plus 0.5em minus 0.4em\relax
  Princeton University Press, 1953, vol.~28, pp. 5--12.

\bibitem{chang2000birkhoff}
C.-S. Chang, W.-J. Chen, and H.-Y. Huang, ``{Birkhoff-von Neumann} input
  buffered crossbar switches,'' in \emph{{IEEE Infocom}}, 2000, pp. 1614--1623.

\bibitem{renganathan2025chronos}
S.~Renganathan and N.~McKeown, ``Chronos: Prescheduled circuit switching for
  {LLM} training,'' in \emph{Proceedings of the 2nd Workshop on Networks for AI
  Computing}, 2025, pp. 89--97.

\bibitem{crouse}
D.~F. Crouse, ``On implementing {2D} rectangular assignment algorithms,''
  \emph{IEEE Transactions on Aerospace and Electronic Systems}, vol.~52, no.~4,
  pp. 1679--1696, 2016.

\bibitem{gandhi2024moetion}
S.~Gandhi and C.~Kozyrakis, ``{MoEtion}: Efficient and reliable checkpointing
  for mixture-of-experts models at scale,'' \emph{arXiv preprint
  arXiv:2412.15411}, 2024.

\bibitem{cohen2026incast}
E.~Cohen, J.~Yallouz, A.~Shpiner, M.~Silberstein, S.~Ratnasamy, and
  I.~Keslassy, ``Incast-free {MoE} rate-based scheduling,'' \emph{arXiv
  preprint arXiv:2607.26340}, 2026.

\end{thebibliography}

\end{document}